\documentclass[12pt]{article}
\usepackage{eurosym}
\usepackage{amssymb}
\usepackage{epsfig}
\usepackage{graphicx}
\usepackage{amsmath}
\usepackage{amsfonts}
\usepackage{amssymb}
\usepackage{float}
\usepackage{subfigure}
\usepackage{float}
\usepackage{amsmath}
\usepackage{amsfonts}
\usepackage{amssymb}

\DeclareGraphicsExtensions{.pdf}
\graphicspath{{Images/}}
\begin{document}

\date{}
\title{\textbf{\Large Meissner-like effect and conductivity of
 power-Maxwell holographic superconductors}}
\author{ \textbf{{\normalsize Doa Hashemi
Asl}$^{1}$} and \textbf{{\normalsize Ahmad Sheykhi}$^{1,2,3} $\thanks{%
asheykhi@shirazu.ac.ir}}\\
$^{1}$ {\normalsize Physics Department and Biruni Observatory,
College of
Sciences,}\\
{\normalsize Shiraz University, Shiraz 71454, Iran}\\
$^{2}$ {\normalsize Research Institute for Astronomy and
Astrophysics of
Maragha (RIAAM),}\\
{\normalsize P.O. Box 55134-441, Maragha, Iran}\\
$^{3}$ {\normalsize Institut f\"{u}r Physik, Universit\"{a}t
Oldenburg, Postfach 2503 D-26111 Oldenburg, Germany}}

\date{}
\maketitle

\begin{abstract}
The full description of a superconductor requires that it has an
infinite DC conductivity (or zero electrical resistivity) as well
as expels the external magnetic fields. Thus, for any holographic
superconductor which is dual to a real superconductor, it is
necessary to examine, simultaneously, these two features based on
the gauge/gravity duality. In this paper, we explore numerically
these two aspects of the higher dimensional holographic
superconductors, in the presence of a Power-Maxwell
electrodynamics as the gauge field. At first, we calculate the
critical temperature, condensation, conductivity, and
superconducting gap, in the absence of magnetic field and disclose
the effects of both power parameter, $s$, as well as the spacetime
dimensions, $d$, on this quantities. Then, we immerse the
superconductor into an external magnetic field, $B$, and observe
that with increasing the magnetic field, the starting point of
condensation occurs at temperature less than the critical
temperature, $T_{c}$, in the absence of magnetic field. This
implies that at a fixed temperature, we can define a critical
magnetic field, above which the critical temperature goes to zero
which is similar to the Meissner effect in superconductor. In
these indications, we also try to show the distinction of the
conformal invariance of the Power-Maxwell Lagrangian that occurs
for $s=d/4$.

\end{abstract}
\newpage \vspace*{0.2cm}
%%%%%%%%%%%%%%%%%%%%%%%%%%%%%%%%%%%%%%%%%%%%%%%%%%%%%%%%%%%%%%%%%%%%%%%%
\section{Introduction}
The gauge/gravity correspondence \cite{Maldacena,Gub1,Witten},
provides an efficient tool to explore strongly coupled phenomena
in the field theory where the perturbational methods are no longer
available. A version of this duality is called
anti-deSitter/conformal field theory (AdS/ CFT) correspondence,
which has been applied for investigating various aspects of
holographic superconductors, using a gravity dual in a $(d+1)$
dimensional bulk for a superconductor which is localized on the
$d$-dimensional boundary of the bulk \cite{Har1,Har2}. AdS/CFT
suggested that an instability for a charged black hole to develop
scalar hair is dual to a superconducting phase transition. The
investigations on the holographic superconductors have arisen a
lot of attentions in the past years and many features of them such
as the critical temperature $T_{c}$, order parameter, critical
exponents, conductivity and their responses to the magnetic field
have been widely explored in the literatures \cite{Gub2,CH,SAH}.
In these studies most of the holographically dual descriptions for
a superconductor are based on a model in which a simple
Einstein-Maxwell theory is coupled to a charged scalar field
\cite{Gary T.H,Gary T.H2, Wang4,P.ZGJZ, RGC2}. The properties of
holographic superconductors for a given black hole depend on the
the electromagnetic field which is coupled to the charged scalar
field and the underlying theory of gravity. In this direction the
studies on the holographic superconductors have been generalized
to other gravity theories such as Gauss-Bonnet gravity
\cite{Wang1,Wang2,RGC3,cui,Ruth,weyl,Gr,Wang3}. Moreover, the
effects of other electrodynamics such as Born-Infeld, Exponential,
Logarithmic on the physical properties of holographic
superconductors were also explored in
\cite{RGC6,SS,GR,SS2,GG,L.P.J.W,JC,JPC}. Conductivity for the
holographic superconductors were calculated in different
dimensions and in the presence of various electrodynamics
\cite{zeng,kono,Doa,Afsoon,bina,mahya,ghora}. Effects of an
external magnetic field on the holographic superconductors were
investigated by employing analytical and numerical methods
\cite{Har2,Nakano,albash,Nakano2,Ge,Roy,Xue,zhao,SS3}. In most
cases, the focus is on the probe limit of the holographic
superconductor where the matter and scalar fields do not back
react on the spacetime metric. One of the important properties of
superconductor, in addition to the infinite DC conductivity, is
their response to an external magnetic field, a phenomena known as
the Meissner effect.

In the present paper we would like to study several aspects of the
higher dimensional holographic superconductors when the gauge
field is in the form of the Power-Maxwell Lagrangian. The
Power-Maxwell electrodynamics is an extension of the Maxwell
electrodynamics which its action has the form \cite{Hass}
\begin{eqnarray}
I_{m}=-\int{d^{d}x\sqrt{-g}(F_{ab }F^{ab })^s}, \label{Act0}
\end{eqnarray}
where $s$ is a positive integer, $F_{ab }=\partial _{a }A_{b
}-\partial _{b }A_{a }$ is the electromagnetic field tensor and
$A_{a }$ is the electromagnetic potential. In arbitrary
dimensions, when the power is $s=d/4$, the Power-Maxwell
Lagrangian becomes invariant under the conformal transformation
($g_{ab}\rightarrow \Omega ^2 g_{ab}$ and $A_{a }\rightarrow A_{a
}$) and the resulting energy-momentum tensor is traceless
\cite{Hass,SheyPM}. It was argued that the theory of conformally
invariant Maxwell field bring rich physics compared to the linear
standard Maxwell field. Applying the analytical Sturm-Liouville
eigenvalue problem and the numerical shooting method in the
background of $d$-dimensional Schwarzschild AdS black hole, the
behavior of the holographic superconductors have been explored
both in the probe limit \cite{PM} and in the presence of
backreaction \cite{PMb}. In the context of Gauss-Bonnet black
holes, analytical and numerical studies on the properties of the
holographic superconductors with Power-Maxwell electrodynamics
have been investigated \cite{SSM,SSM2}. Using the analytical
matching method, the upper critical magnetic field of holographic
superconductor with conformally invariant Power-Maxwell
electrodynamics have been explored in \cite{SSD}. It was argued
that in the presence of magnetic field, the physically acceptable
phase behavior of the holographic superconductor can be deduced
only for $s={d}/{4}$, which guaranties the conformal invariance of
the Power-Maxwell Lagrangian. Till now, conductivity as well as
the effects of the magnetic field on the properties of the higher
dimensional Power-Maxwell holographic superconductors have not
been explored by using the numerical shooting method. The novelty
of our work comes from the fact that we employ the numerical
method for investigating the Meissner-like effects, while most
studies on the effect of magnetic field is based on the analytical
methods \cite{SSD}. The purpose is to see whether, when the gauge
field is in the Power-Maxwell form, the physically acceptable
phase behavior of the holographic superconductor in the presence
of magnetic field can be deduced only for the conformal invariance
power parameter, $s=d/4$, as argued in \cite{SSD}, or it may admit
reasonable behavior for other values of the power parameter $s$.

Our aim in this paper is to investigate the responses of the
Power-Maxwell holographic superconductors on the electric field as
well as on the external magnetic field in all higher dimensions.
As we expect the properties of the holographic superconductor is
affected by the choice of the electrodynamics. We shall do our
calculations for several values of the power parameter and also
compare the results for the case of conformal invariant with other
cases. Throughout this paper, we neglect the back reaction of the
scalar field and the gauge field for simplicity, so we cannot see
all of the signature physics of a superconductor. However, in the
final step, in order to examine the full effects of the external
magnetic field on the holographic superconductor, we consider the
the effects of backreaction of the magnetic field on the
background geometry.

The rest of the paper is organized as follows. In the next
section, we introduce the basic field equations of the
$d$-dimensional holographic superconductor with Power-Maxwell
electrodynamics and employ the numerical shooting method to obtain
a relation between the critical temperature $T_{c}$ and charge
density $\rho$. We also calculate the values of the condensation
operator and the critical exponent in this section. In section
\ref{Cond} we study the holographic electrical conductivity of the
system and reveal the response of the system to an external
electric field. Finally, in the last section \ref{magnetic}, we
study the effects of an external magnetic field and observe the
Meissner-like effect.
%%%%%%%%%%%%%%%%%%%%%%%%%%%%%%%%%%%%%%%%%%%%%%%%%%%%%%%%%%%%%%%%%%%%%%%%%%%%%%%%%%%%%%%%%%%%%%%%%%%%%%%
\section{ Higher dimensional holographic superconductor with Power-Maxwell electrodynamics} \label{basic}
\subsection{ The background}
We begin with the higher dimensional Einstein action coupled to a
massive charged complex scalar field in the presence of
Power-Maxwell field with a negative cosmological constant,
\begin{equation}\label{action}
S=\int d^{d}x\sqrt{-g}\left[ R-2\Lambda +\beta(-\mathcal{F})^{s}-|\nabla
\psi -iqA\psi |^{2}-m^{2}|\psi |^{2}\right] ,
\end{equation}%
where $R$ is the Ricci scalar, and the $d$-dimensional
cosmological constant is defined as $\Lambda
=-{(d-1)(d-2)}/{2l^{2}}$, where $l$ is the AdS radius
\cite{Alfonso V.R}. Here $\beta$ is a constant, $
\mathcal{F}=F_{ab }F^{ab }$ where $A_{a}$ and $\psi$ are
respectively, gauge field and scalar field with charge $q$ and
mass $m$. The Power-Maxwell contribution to the action is given by
$\beta(-\mathcal{F})^{s}$ with the power parameter $s$. When $s=1$
and $\beta=1/4$ it reduces to the Maxwell case. In this paper,
without loss of generality, we keep $\beta=1/4$ and only change
the power parameter $s$.

Varying action (\ref{action}) with respect to the metric $g_{a
b}$, the gauge field $A_a$ and the scalar field $\psi$ yields the
following field equations
\begin{eqnarray}\label{enequation}
&&R_{ac}-\frac{g_{ac} R}{2}+\Lambda g_{ac}=\frac{1}{8}g_{ac}
 (-\mathcal{F})^{s}
+\frac{s}{2}  (-\mathcal{F})^{s-1} F_{ad} F_{c}^{\ d}-\frac{g_{ac}}{2}~m^2 |\psi|^2 \nonumber\\
 &&-\frac{g_{ac}}{2}
|\nabla\psi-iqA\psi|^2+\frac{1}{2}\left[(\nabla_{a}\psi-iqA_{a}\psi)(\nabla_{c}\psi^{*}+iqA_{c}\psi^{*})+a\leftrightarrow
c\right]~,
\end{eqnarray}
\begin{eqnarray}\label{gaugefield}
s \nabla^{a}\left[F_{ac}(-\mathcal{F})^{s-1}
\right]=iq\left[\psi^{*}(\nabla_{c}-iqA_{c})\psi-\psi(\nabla_{c}+iqA_{c})\psi^{*}\right]~,
\end{eqnarray}
\begin{eqnarray}\label{scalarfield}
(\nabla_{a}-iqA_{a})(\nabla^{a}-iqA^{a})\psi-m^2\psi=0.
\end{eqnarray}
In this section, we consider the limit where the scalar field
$\psi$ and  the gauge field $A_{a}$ do not back-react on the
background geometry. Thus the solution for the background geometry
we take is the $d$-dimensional AdS-Schwarzschild black hole which
is described by the following metric
\begin{equation}\label{metric}
ds^2=-f(r)dt^2+\frac{1}{f(r)}dr^2+ r^2 h_{ij} dx^{i} dx^{j},
\end{equation}
where $f(r)$ is given by
\begin{eqnarray}\label{fequation}
f(r)=\frac{r^{2}}{l^{2}}-\frac{1}{r^{d-3}}\left(\frac{r^{d-1}_{+}}{l^{2}}
\right),
\end{eqnarray}
and $r_{+}$ is the event horizon radius. As expected, by
neglecting the backreaction of the matter fields on the background
geometry, just the existence of the black hole and the AdS radius
are sufficient for determining the metric function. The Hawking
temperature, which will be interpreted as the temperature of the
holographic superconductors, is given by
\begin{eqnarray} \label{T}
T=\frac{f^{\prime}(r_{+})}{4\pi}=\frac{(d-1)r_{+}}{4\pi l^{2}}.
\end{eqnarray}
In order to simplify the field equations and following
\cite{Har1}, we choose a particular form for the gauge field
$A_{a}$ and scalar field $\psi$ which is based on the fact that we
know that what is the minimum content needed to produce a
gravitational dual to a superconductor. In this step, that we want
to explain exclusively a superconductor that exhibits condensation
below the critical temperature $T_{c}$ (and without any external
electric or magnetic field), that is enough to turn on only
$A_{t}$ component by introducing the scalar hair for the black
hole as \cite{Har1},
\begin{eqnarray}\label{Aandpsi}
A_{a}dx^{a}=\phi(r)~dt, \  \   ~~\psi=\psi(r),
\end{eqnarray}
which depend only on $r$ coordinate. Inserting metric
(\ref{metric}), scalar and gauge fields (\ref{Aandpsi}) in the
field equations (\ref{gaugefield}) and (\ref{scalarfield}), we can
arrive at the following pair of coupled second order differential
equations for the gauge and scalar fields
\begin{eqnarray}  \label{phir}
\phi^{\prime\prime}(r)
-\frac{d-2}{(1-2s)r} \phi^{\prime}(r) +\frac{q^{2} \psi^{2} (r) \phi^{\prime (2-2s)}(r) }
{ s 2^{s-2} (1-2s) f(r)}\phi(r)=0,  \label{Eqphi}
\end{eqnarray}
\begin{eqnarray}  \label{psir}
\psi^{\prime\prime}(r) +\left(\frac{f^{\prime
}}{f}+\frac{d-2}{r}\right) \psi^{\prime }(r) +\bigg(\frac
{q^{2}\phi^2(r)}{f^2(r)}-\frac{m^2}{f(r)}\bigg) \psi(r)=0,
\end{eqnarray}
with appropriate boundary conditions at the horizon and at the
asymptotic infinity. We require at the horizon $(r=r_{+})$,
\begin{eqnarray}\label{boundrplus}
\phi(r_{+})=0,~~~
\psi(r_{+})=\frac{f^\prime(r_{+})\psi^\prime(r_{+})}{m^{2}}.
\end{eqnarray}
While at the boundary ($r\rightarrow \infty$), the solutions
behave like
\begin{eqnarray}  \label{bound1}
\phi(r)&=&\mu-\frac{\rho^{\frac{1}{2s-1}}}{r^{\frac{d-2}{2s-1}-1}},\\
\psi(r)&=&\frac{\psi_{-}}{r^{\Delta_{-}}}+\frac{\psi_{+}}{r^{\Delta_{+}}},
\label{bound2}
\end{eqnarray}
where $\mu$ and $\rho$ are, respectively, the chemical potential
and charge density of dual field theory, and
\begin{eqnarray}\label{delta}
\Delta_{\pm}=\frac{1}{2}\left[(d-1)\pm\sqrt{(d-1)^2+4m^2
l^2}\right].
\end{eqnarray}
In contrast to other electrodynamics, in case of the Power-Maxwell
electrodynamics, the asymptotic behavior of $\phi$ depends on the
electrodynamics parameter $s$, and since the gauge field should
have a finite value at the boundary $(r\rightarrow\infty)$, we may
arrive a restriction on the value of $s$. The meaningful range of
the parameter $s$ is obtained $1/2<s<(d-1)/2$ \cite{SSM}, thus at
any dimension, we make our choices for power parameter with
respect to this range.

The coefficients $\psi_{-}$ and $\psi_{+}$ both multiply
normalizable modes of the scalar field equations and according to
the AdS/CFT correspondence, they correspond to the expectation
value of condensation operator on the boundary \cite{Har1}.
Indeed, either $\psi_{+}$ or $\psi_{-}$ can be dual to the value
of the condensation operator, and the other one is dual to its
source. Following \cite{Gary T.H}, we can impose the boundary
condition in which either $\psi_{-}$ or $\psi_{+}$ vanishes. In
this paper we set $\psi_{-}=0$. One can solve the field equations
(\ref{phir}) and (\ref{psir}) by different methods. The method we
choose for this purpose is a numerical shooting method. At first,
we make the coordinate transformation as $r\rightarrow
z={r_{+}}/{r}$ in the equations and work with dimensionless
parameter $z$. We also assume the values of $q$, $r_{+}$ and $l$
to be unity according to the symmetries. In addition, we make our
choices for the values of the parameters, in the allowed ranges.
In this regards, $m^2$ must satisfy the BF
(Breitenlohner-Freedman) bound \cite{Alfonso V.R}, which in our
case is given by $m^{2}\geq -{(d-1)^{2}}/{4l^{2}}$.
%%%%%%%%%%%%%%%%%%%%%%%%%%%%%%%%%%%%%%%%%%%%%%%%%%%%%%%%%%%%%%%%%%%%%%%%%%%%%%%%%%%%%%%%%%%%%%%%%%%%%
\subsection{Critical temperature} \label{Crit}
Let us start by deriving the critical temperature, $T_{c}$, of the
higher dimensional holographic superconductor in the presence of
Power-Maxwell electrodynamics. Our investigation will be followed
by the method  developed in \cite{Doa}. In this calculation we
also use the charge density $\rho$ to fix a scale. From QFT
dimensional analysis, for the Power-Maxwell Lagrangian in a
$d$-dimensional AdS bulk, we find that the dimension of any
component of gauge field $A_{a}$ is always one, independent of the
power parameter $s$. Thus, from Eq. (\ref{bound1}), the dimension
of charge density is obtained $(d-2)$, regardless of the value of
the power parameter $s$. Since the dimension of the temperature is
one, therefore in order to have the dimensionless quantities, one
should consider $\rho^{1/(d-2)}$. In tables \ref{tabale1}, we
summarize the results of the critical temperature for different
values of $s$ and $d$ with boundary condition $\psi_{-}=0$ and the
mass of scalar field $m^2=-2$,$-3$ and $-4$ respectively for
$d=4$, $5$ and $6$.

\begin{table}[tbh]
    \label{tabale1} \centering
    \begin{tabular}{cccccc}
        \hline\hline $s$   & $T_{c}(d=4)$   & $T_{c}(d=5)$   &
        $T_{c}(d=6)$    &  \\%
        [0.05ex] \hline
        $3/4$   &   0.194 $\rho^{1/2}$ &  - &  -    \\
        $1$ &  0.118 $\rho^{1/2}$ &   0.198 $\rho^{1/3}$ &   0.271$\rho^{1/4}$  \\
        $5/4$ &  0.045 $\rho^{1/2}$ &    0.128 $\rho^{1/3}$ &   0.202$\rho^{1/4}$   \\
        $6/4$ &  - &    0.068 $\rho^{1/3}$ &   0.138$\rho^{1/4}$  \\
        $7/4$ &   - & - &    0.085$\rho^{1/4}$  \\
        [0.5ex] \hline
    \end{tabular}%
    \caption{Critical temperature $T_{c}$ for $d=4$, $5$ and $6$ with
        $\Delta_{+}=2$, $3$ and $4$.}\label{tabale1}
\end{table}
From this table, we find that with increasing the power parameter
$s$, the critical temperature will be decreased as well, which
implies that the condensation is harder to form.  Moreover, the
results are also consistent with the general understanding that
the phase transition is easier to achieve in higher dimensional
systems. For completeness, in this table, besides the Maxwell case
where $s=1$, in any dimension $d$, we choose three values for the
power parameter $s$, namely the conformal invariance case,
$s=d/4$, the case $s<d/4$ and the case $s>d/4$.
%%%%%%%%%%%%%%%%%%%%%%%%%%%%%%%%%%%%%%%%%%%%%%%%%%%%%%%%%%%%%%%%%%%%%%%%%%%%%%%%%%%%
\subsection{Condensation operator and critical exponent}
The first expected result in
the study of a holographic superconductor is  quickly increase of
the order parameter by reaching critical temperature $T_{c}$. Here
the order parameter is a charged operator in the boundary that is
related by AdS/CFT to the coefficients of the asymptotic behavior
of the scalar field according to Eq.(\ref{bound2}), which based on
our choice ($\psi_{-}=0$), $\psi_{+}$ gives us the expectation
value of the condensation operator \cite{Har2}. Solving the field
equations, we obtain $\psi_{+}$ (and thus
$\langle\mathcal{O_{+}}\rangle$) values. Using the Lagrangian
dimensional analysis, we obtain $\Delta{+}$ as the dimension of
$\psi_{+}$, and hence the dimension of condensation operator will
be the same. So it is easy to make dimensionless parameters for
the plotting. we plot for three dimensions of spacetime: $d=4$,
$5$ and $6$ and at any dimension for three values of power
parameter $s$.
\begin{figure}[H]
    \begin{center}
        \begin{minipage}[b]{0.325\textwidth}\begin{center}
                \subfigure[~$d=4$]{
                    \label{fig2a}\includegraphics[width=\textwidth]{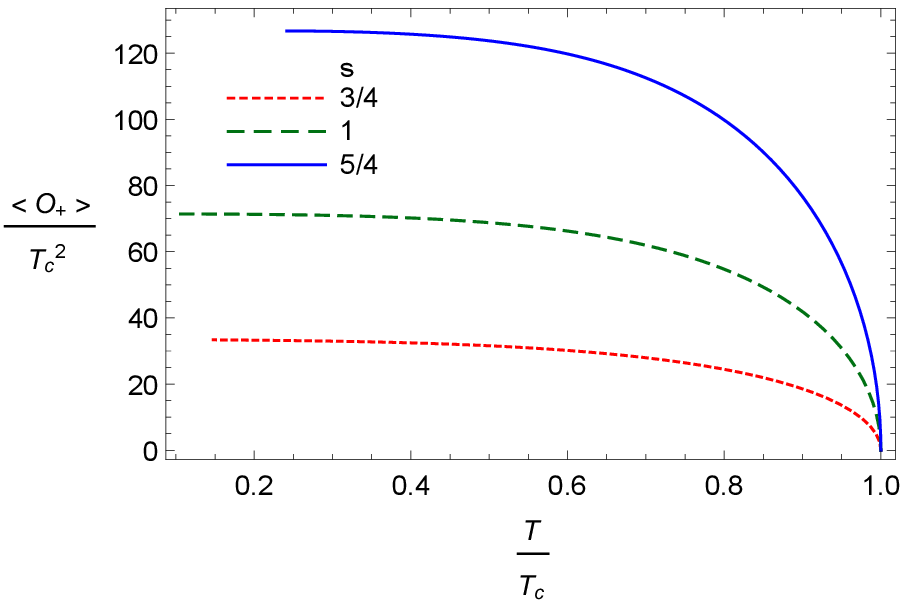}\qquad}
        \end{center}\end{minipage} \hskip+0cm
        \begin{minipage}[b]{0.325\textwidth}\begin{center}
                \subfigure[~$d=5$]{
                    \label{fig2b}\includegraphics[width=\textwidth]{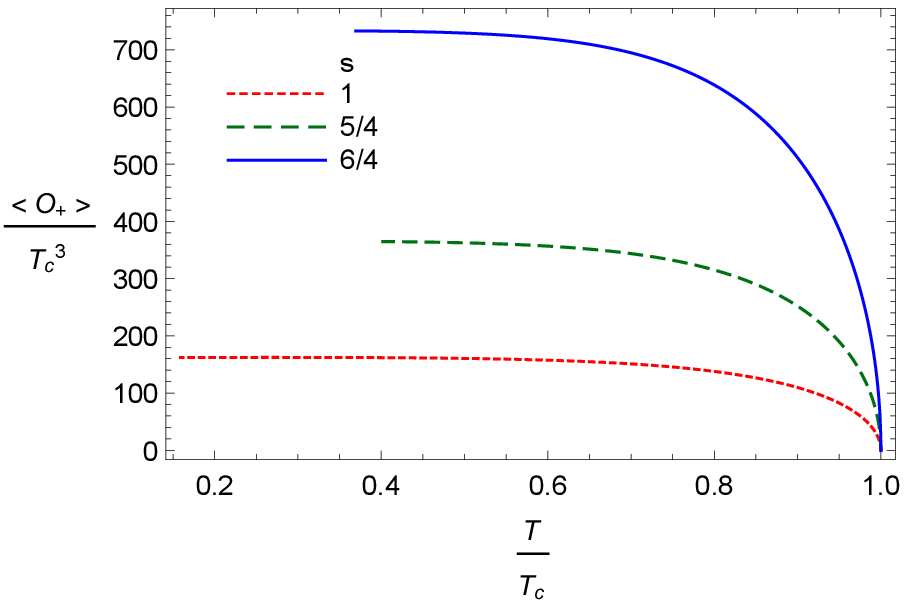}\qquad}
        \end{center}\end{minipage} \hskip0cm
        \begin{minipage}[b]{0.325\textwidth}\begin{center}
                \subfigure[~$d=6$]{
                    \label{fig2c}\includegraphics[width=\textwidth]{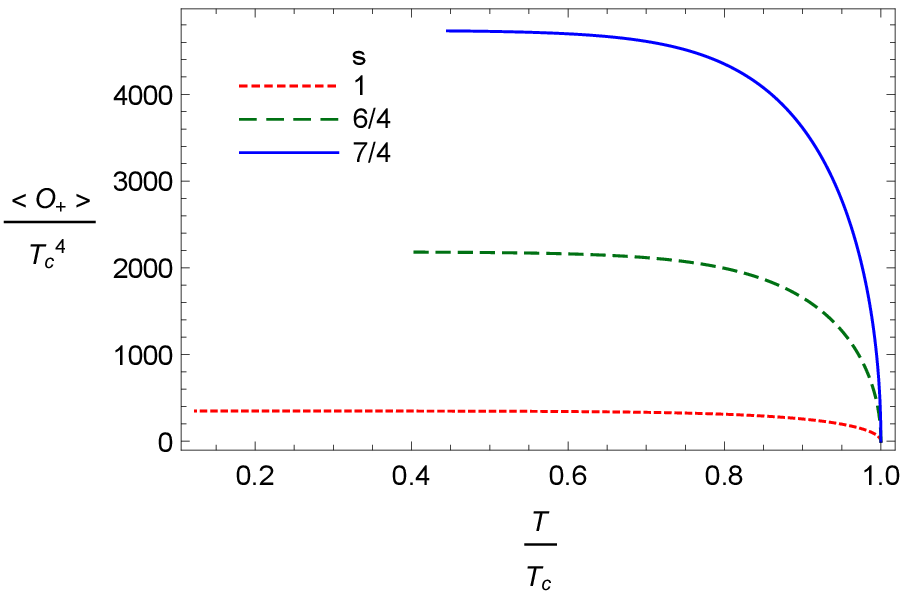}\qquad}
\end{center}\end{minipage} \hskip0cm
\end{center}
\caption{Dimensionless condensation operator
$\left(\frac{\langle\mathcal{O_{+}}\rangle}{T_{c}^{\Delta_{+}}}\right)$
versus dimensionless temperature $T/T_{c}$ for $d=4$, $5$ and $6$.
} \label{fig2}
\end{figure}
As seen from the figures, the values of the order parameter
becomes nonzero at $T=T_{c}$ and grow up quickly when
$T\rightarrow0$. This means the charged scalar operator has
condensed. Moreover, from the theory of superconductor
\cite{cooper}, near the critical temperature ($T\sim T_{c}$), as
the critical temperature approaches from below, we find the order
parameter as
\begin{eqnarray}\label{orderparameter}
\langle\mathcal{O_{+}}\rangle=\gamma T_{c}^{\Delta_{+}}\left(1-\frac{T}{T_c}\right)^{%
    {1}/{2}},
\end{eqnarray}
where $\gamma$ is a constant and $1/2$ is the critical exponent
for the second order phase transition. Also from analytical
methods (Sturm-Liovile or matching method), this result was
obtained in earlier studies, and except the coefficient $\gamma$,
the form of this relation is in a general form that do not depend
on the choice of the electrodynamics and the dimension of the
spacetime. Therefore, it is interesting to fit the plots we
obtained from the numerical shooting method for the holographic
superconductor, on the formula like (\ref{orderparameter}) and
calculate the coefficient $\gamma$ and critical exponent. The
results are summarized in table \ref{tabale2}.
\begin{table}[H]
    \label{tabale2} \centering
    \begin{tabular}{cccccc}
        \hline\hline $s$   & $\sqrt{2}\gamma (d=4)$   & $\gamma (d=5)$   &
        $\gamma (d=6)$    &  \\%
        [0.05ex] \hline
        $3/4$   &   57.3 &  - &  -    \\
        $1$ &  137.3 &  384.4 &  942.2  \\
        $5/4$ &  255.5 &    896.3 &   2593.5   \\
        $6/4$ &  - &    1830.4 &   6199.9  \\
        $7/4$ &   - & - &   13549.2  \\
        [0.5ex] \hline
    \end{tabular}%
    \caption{The values of
        $\protect\gamma$ for $d=4$,$5$ and $6$.}\label{tabale2}
\end{table}
Besides, for all values of the power parameter $s$ and $d$, we
obtain a value of about 1/2 for the critical exponent. From table.
\ref{tabale2} and Fig. \ref{fig2}, we can see that in higher
dimensions and for larger value of $s$, $\gamma$  and (as a
result) the rescaled condensation operator
($\frac{\langle\mathcal{O_{+}}\rangle}{T_{c}^{\Delta_{+}}}$) have
explicitly increase, that arises from the fact that in these cases
the critical temperature decreases. Until now, we have introduced
and built a model of holographic superconductor in the presence of
the power-Maxwell electrodynamics and observed the condensation
for this kind of holographic superconductor. However, as we
pointed out earlier, this study is not enough and the model can
describe a real dual to superconductor if displays both properties
of the infinite DC conductivity as well as expelling the magnetic
field lines in the superconducting phase. So from now on, in order
to study these two properties of this superconductor, in the first
step, we apply an external electric field and in the second step,
we apply an external magnetic field. In other words, we will
examine the effects of background electric and magnetic fields on
the holographic superconductor in the presence of the
power-Maxwell electrodynamics. As we shall see the presence of
this electrodynamics affects the conductivity and the way of
expelling the applied magnetic field.
%%%%%%%%%%%%%%%%%%%%%%%%%%%%%%%%%%%%%%%%%%%%%%%%%%%%%%%%%%%%%%%%%%%%%%%%%%%%%%%%%%%%%%%%%%%%
\section{Conductivity} \label{Cond}
In this section, we will apply an electric field in order to study
the transport phenomena in the holographic superconductors with
power-Maxwell field. Since the purpose is to obtain the DC and AC
conductivity (that means conductivity in different frequency), we
should consider the applied field to fluctuate relative to time
with the frequency $\omega$ in the boundary. The result of
illustrating the above concepts by employing AdS/CFT
correspondence is considering fluctuation of the field $A_{x}$ in
the bulk. This fluctuation is dual to the electric current $J^{x}$
operator in the CFT \cite{Har1,Har2}. We take all the currents and
sources in the $x$ direction. It should be noted that if we
consider the backreaction of the matter fields on the spacetime
geometry, we should also take into account the fluctuation of the
$g_{tx}$ component of the metric. For simplicity, here we neglect
the backreaction. Thus, we assume the Maxwell field as
 \begin{eqnarray}\label{Aacond}
 A_{a}dx^{a}=\phi(r)~dt+A_{x}(r)e^{-i\omega t} dx.
 \end{eqnarray}
Our strategy is to obtain the field equation of $A_{x}(r)$ and try
to solve it with appropriate boundary conditions in the bulk for
different values of the power parameter and dimension of
spacetime. Then, we  shall use this solution to find out the
current as well as the conductivity in the boundary. From
Eq.(\ref{gaugefield}) the field equation for $A_{x}(r)$ is given
by
\begin{eqnarray}\label{key}
 &&A_{x}^{\prime \prime }(r)+\bigg(\frac{2(s-1)\phi ^{\prime \prime }(r)}{\phi
 ^{\prime }(r)}+ \frac{(d-4)}{r}+\frac{f^{\prime }(r)}{f(r)}\bigg)A_{x}^{\prime }(r)
\\ \notag &&+\bigg(\frac{\omega ^{2}}{f^{2}(r)}-\frac{\psi^{2}(r)}{
  2^{(s-2)}~s~ \phi^{\prime (2s-2)}(r) f(r)}\bigg)A_{x}(r)=0.
 \end{eqnarray}%
It should be noted that in this calculation we linearize the
equation that means working in the regime of linear response. It
is a matter of calculations to show that the asymptotic behavior
of this equation for $r\rightarrow\infty$ is given by
\begin{equation*}
A_{x}^{\prime \prime }(r)+\frac{d-2}{(1-2s)r} A_{x}^{\prime
 }(r)+\frac{\omega^{2}}{r^4} A_{x}(r)=0.
\end{equation*}
It is apparent that the solutions of this equation depend on the
both $s$ and $d$ parameters. Therefore, we present the solution
for several values of $d$ and $s$. The solution for $d=4$ is:
\begin{equation}\label{aaad4}
A_{x}=\left\{
\begin{array}{ll}
A^{(0)}+\frac{\omega ^{2}A^{(0)}}{2r^{2}}+\frac{A^{(1)}}{r^3}+...\ , & \ \ ~~%
\mathrm{for}\ s=3/4, \\
& \\
A^{(0)}+\frac{A^{(1)}}{r}-\frac{\omega ^{2}A^{(0)}}{2r^{2}}+..., &
\ \quad \mathrm{for}\ s=1, \\
& \\
A^{(0)}+\frac{A^{(1)}}{r^{1/3}}-\frac{3 \omega ^{2}A^{(0)}}{10r^{2}}+\frac{3 \omega^{2} A^{(1)}}{14r^{7/3}}+...\ , & \ \ ~~%
\mathrm{for}\ s=5/4. \\
& \\
\end{array}%
\right.
\end{equation}%
For $d=5$:
\begin{equation}\label{aaad5}
A_{x}=\left\{
\begin{array}{ll}
A^{(0)}+\frac{A^{(1)}}{r^{2}}+\frac{\omega ^{2}A^{(0)}ln(kr)}{2r^{2}}+...\ , & \ \ ~~%
\mathrm{for}\ s=1, \\
& \\
A^{(0)}+\frac{A^{(1)}}{r}-\frac{\omega ^{2}A^{(0)}}{2r^{2}}-\frac{ \omega^{2} A^{(1)}}{6 r^3}+..., &
\ \quad \mathrm{for}\ s=5/4, \\
& \\
A^{(0)}+\frac{A^{(1)}}{r^{1/2}}-\frac{ \omega ^{2}A^{(0)}}{3r^{2}}-\frac{\omega^{2} A^{(1)}}{5r^{10/4}}+...\ , & \ \ ~~%
\mathrm{for}\ s=6/4. \\
& \\
\end{array}%
\right.
\end{equation}%
And for $d=6$:
\begin{equation}\label{aaad6}
A_{x}=\left\{
\begin{array}{ll}
A^{(0)}+\frac{\omega ^{2}A^{(0)}}{2r^{2}}+\frac{A^{(1)}}{r^{3}}-\frac{\omega ^{4}A^{(0)}}{8r^{4}}-\frac{A^{(1)} \omega^{2}}{10r^{5}}+...\ , & \ \ ~~%
\mathrm{for}\ s=1, \\
& \\
A^{(0)}+\frac{A^{(1)}}{r}-\frac{\omega ^{2}A^{(0)}}{2r^{2}}-\frac{ \omega^{2} A^{(1)}}{6 r^3}+\frac{\omega ^{4}A^{(0)}}{24r^{4}}-\frac{ \omega^{4} A^{(1)}}{120 r^5}-\frac{\omega ^{6}A^{(0)}}{720r^{6}}+..., &
\ \quad \mathrm{for}\ s=6/4, \\
& \\
A^{(0)}+\frac{A^{(1)}}{r^{3/5}}-\frac{5 \omega ^{2}A^{(0)}}{14r^{2}}-\frac{5 \omega^{2} A^{(1)}}{26r^{13/5}}+\frac{25 \omega ^{4}A^{(0)}}{925r^{4}}+\frac{25 \omega^{4} A^{(1)}}{2392r^{23/5}}+...\ , & \ \ ~~%
\mathrm{for}\ s=7/4, \\
& \\
\end{array}%
\right.
\end{equation}%
where $A^{(0)}, A^{(1)}$ and $k$ are constant parameters.
Following Ref. \cite{D.Tong}, for calculating the boundary current
operator, we first need the on shell action $S_{o.s.}$. Then one
can use the following relation for calculating the current,
\begin{equation}
J=\frac{\delta S_{o.s}}{\delta A^{(0)}}.  \label{current}
\end{equation}
The on shell action is given by \cite{D.Tong}
\begin{equation}
S_{o.s}=\int_{r_{+}}^{\infty }dr\int
d^{d-1}x\sqrt{-g}\mathcal{L},
\end{equation}
where in our case it reduces     to
\begin{equation}\label{onshellaction}
S_{o.s}=\int d^{d-1}x~\bigg(-2^{(s-2)}~s~r^{d-4}f(r)~\phi^{\prime (2s-2)} A_{x}(r)A_{x}^{\prime }(r)\bigg)\big|_{r\rightarrow \infty }.
\end{equation}
Combining Eqs.(\ref{aaad4}), (\ref{aaad5}), (\ref{aaad6}) and
(\ref{bound1}) with  Eq.(\ref{onshellaction}), we obtain
explicitly the on shell action for different values of the power
parameter $s$ and for different spacetime dimension $d$. Then,
employing Eq.(\ref{current}), we  can find the current for
different values of $s$  and $d$. But most probably in this
calculation, we may encounter with divergence terms in the action
when we set $r\rightarrow\infty$. In order to remove the
divergences in the action, we add a proper counter term
\cite{skenderis}. We need counter terms for $d=4$ and $s=3/4$,
$d=5$ and $s=1$, $d=6$ and $s=1$. For the economic reasons we do
not present here the counter terms. One can calculate the
conductivity for the system from $\sigma _{xx}={J_{x}}/{E_{x}}$,
by having the applied electric field and the current caused by it.
The electric field is $E_{x}=-\partial _{t}A_{x}=i \omega A^{(0)}
$. Thus, for the conductivity we find out for $d=4$,
\begin{equation}
\sigma =\left\{
\begin{array}{ll}
\frac{3 \sqrt{3} A^{(1)}}{4*2^{1/4}~i\omega \rho~A^{(0)}}\ , & \ \ ~~\mathrm{for}\ s=3/4,\\
&  \\
\frac{A^{(1)}}{i\omega A^{(0)}}, & \ \quad
\mathrm{for}\ s=1,\\
&  \\
\frac{5A^{(1)} ~\rho^{1/3}}{6\sqrt{3}*2^{3/4}~i\omega A^{(0)}}\ , & \ \ ~~\mathrm{for}\ s=5/4.\\
&  \\
\end{array}%
\right.  \label{condd4}
\end{equation}%
For $d=5$ we find
\begin{equation}
\sigma =\left\{
\begin{array}{ll}
\frac{2A^{(1)}}{i\omega A^{(0)}}+\frac{i\omega }{2}\ , & \ \ ~~\mathrm{for}\ s=1,\\
&  \\
\frac{5A^{(1)}\rho^{1/3}}{2*2^{3/4}~i\omega A^{(0)}}, & \ \quad
\mathrm{for}\ s=5/4,\\
&  \\
\frac{3A^{(1)} \rho^{1/2}}{4\sqrt{2}~i\omega A^{(0)}}\ , & \ \ ~~\mathrm{for}\ s=6/4.\\
&  \\
\end{array}%
\right.  \label{condd5}
\end{equation}%
Finally, for $d=6$, we obtain
\begin{equation}
\sigma =\left\{
\begin{array}{ll}
\frac{3A^{(1)}}{i\omega A^{(0)}}\ , & \ \ ~~\mathrm{for}\ s=1,\\
&  \\
\frac{3A^{(1)} \rho^{1/2}}{\sqrt{2}i\omega A^{(0)}}, & \ \quad
\mathrm{for}\ s=6/4,\\
&  \\
\frac{63\sqrt{3}~A^{(1)}\rho^{3/5}}{50\sqrt{5}*2^{1/4}i\omega A^{(0)}}\ , & \ \ ~~\mathrm{for}\ s=7/4.\\
&  \\
\end{array}%
\right.  \label{condd6}
\end{equation}%
Let us note that the relations for $\sigma$, contain a power of
$\rho$ for some values of $s$. This comes from the fact that
Eq.(\ref{onshellaction}), includes $\phi^{\prime (2s-2)}$. Using
Eq.(\ref{bound1}) for $\phi$, which is a function of $\rho$ and
$s$, the dependence of $\sigma$ on the charge density $\rho$ is
expected. Next, we solve numerically the Eq.(\ref{key}) in the
bulk. Having at hand the behavior of the field $A_{x}(r)$ in the
bulk, one can find its behavior at the boundary, that determines
the values of $A^{(0)}$ and $A^{(1)}$. Thus, we can calculate the
conductivity as a function of frequency $\omega$, for all
supposing values of $d$ and $s$ from Eq. (\ref{condd4}),
(\ref{condd5}) and (\ref{condd6}).\\
 \begin{figure*}[h!]
    \begin{center}
        \begin{minipage}[b]{0.325\textwidth}\begin{center}
                \subfigure[ ~$d=4,~s=3/4$]{
                    \label{fig5a}\includegraphics[width=\textwidth]{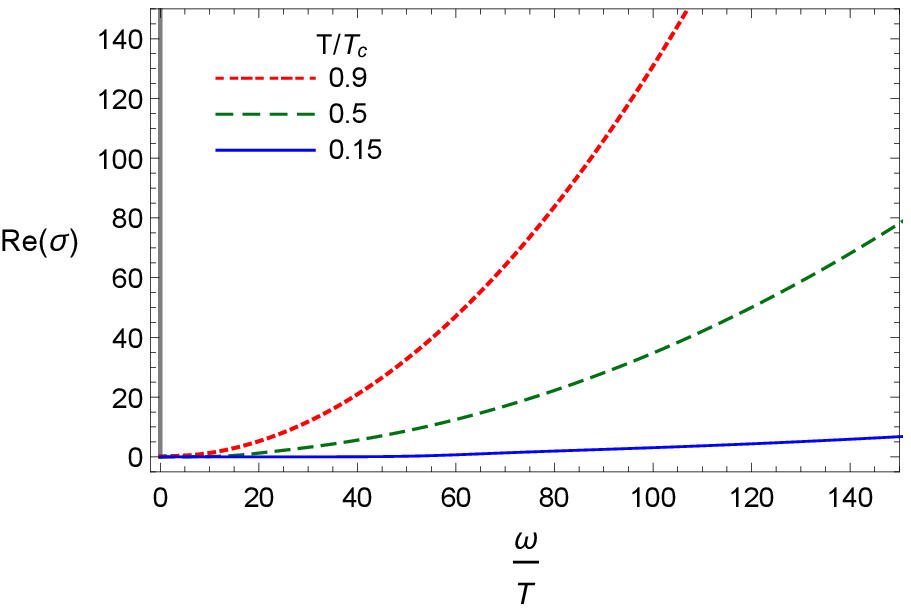}\qquad}
        \end{center}\end{minipage} \hskip+0cm
        \begin{minipage}[b]{0.325\textwidth}\begin{center}
                \subfigure[ ~$d=4, ~s=1$]{
                    \label{fig5b}\includegraphics[width=\textwidth]{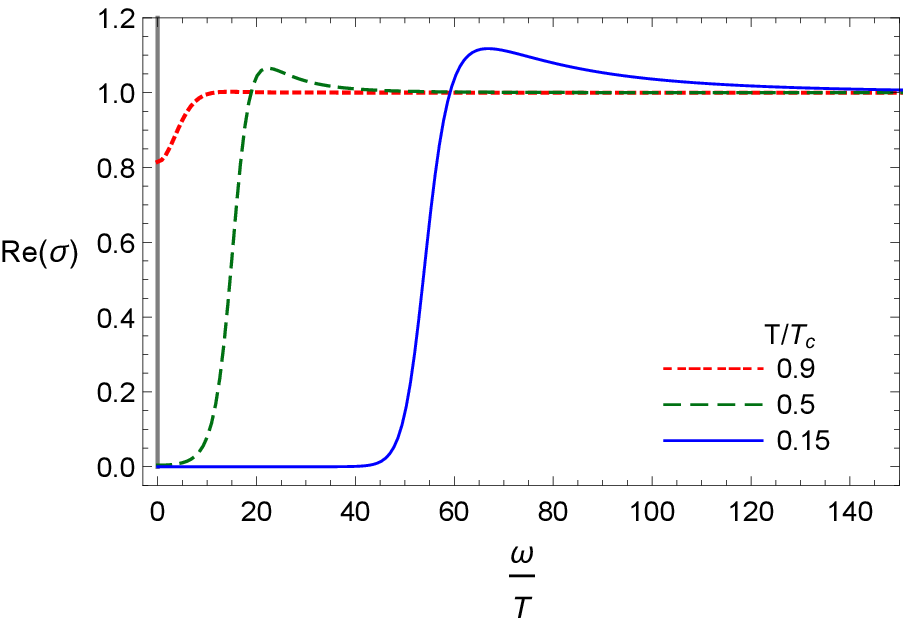}\qquad}
        \end{center}\end{minipage} \hskip0cm
        \begin{minipage}[b]{0.325\textwidth}\begin{center}
                \subfigure[ ~$d=4,~s=5/4$]{
                    \label{fig5c}\includegraphics[width=\textwidth]{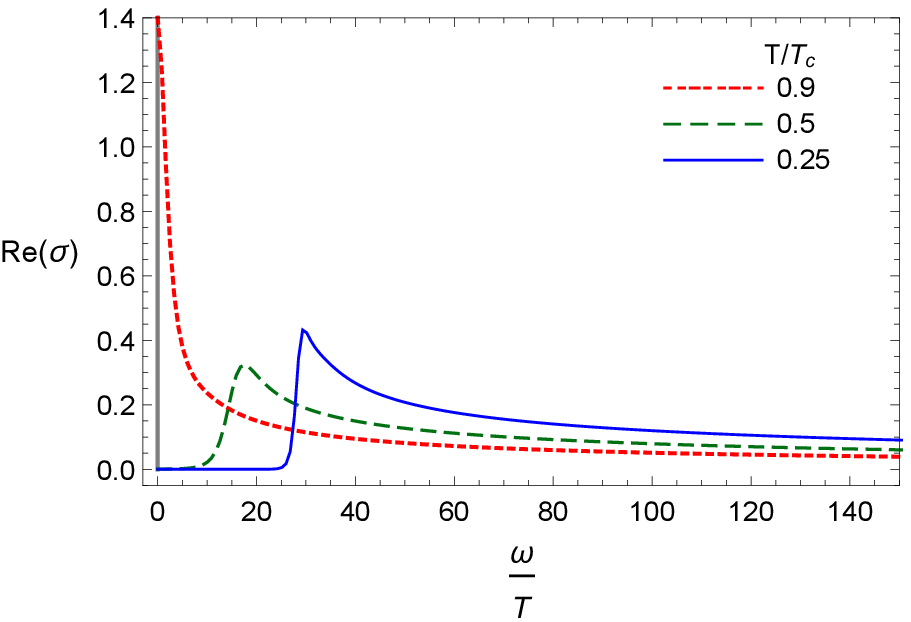}\qquad}
        \end{center}\end{minipage} \hskip0cm
        \begin{minipage}[b]{0.325\textwidth}\begin{center}
                \subfigure[$~ d=5,~s=1$]{
                    \label{fig5d}\includegraphics[width=\textwidth]{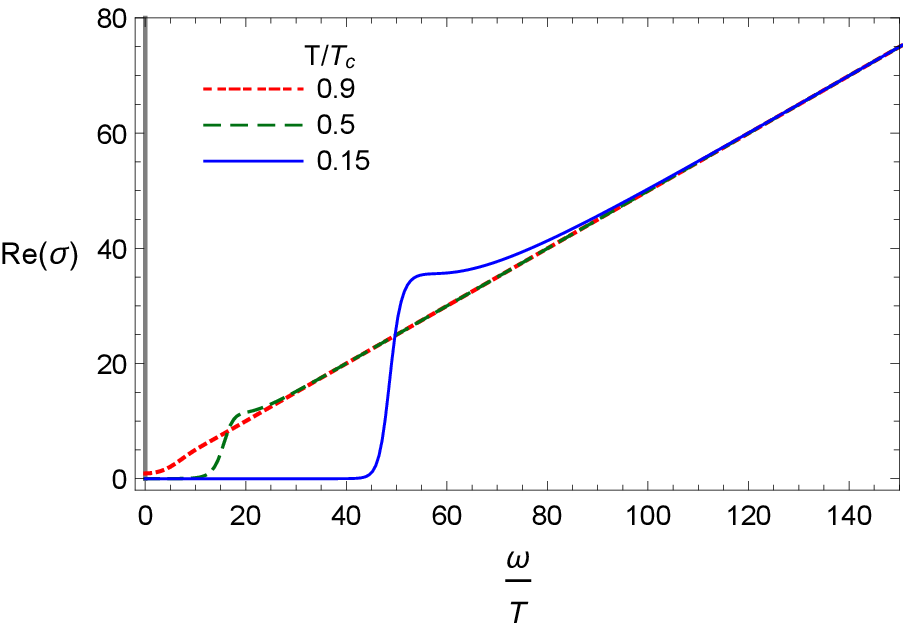}\qquad}
        \end{center}\end{minipage} \hskip0cm
        \begin{minipage}[b]{0.325\textwidth}\begin{center}
                \subfigure[$~ d=5,~s=5/4$]{
                    \label{fig5e}\includegraphics[width=\textwidth]{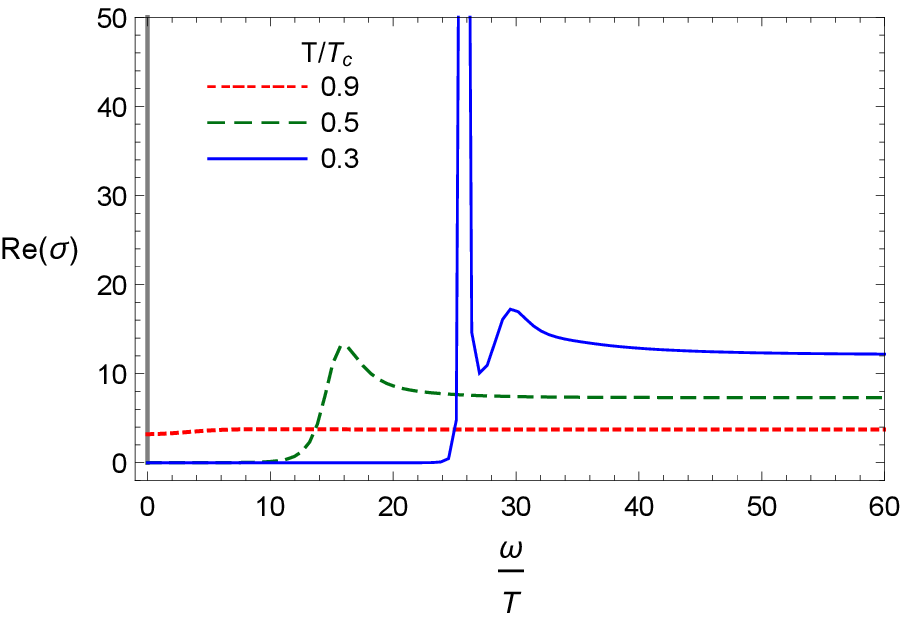}\qquad}
        \end{center}\end{minipage} \hskip0cm
        \begin{minipage}[b]{0.325\textwidth}\begin{center}
                \subfigure[$~ d=5,~s=6/4$]{
                    \label{fig55f}\includegraphics[width=\textwidth]{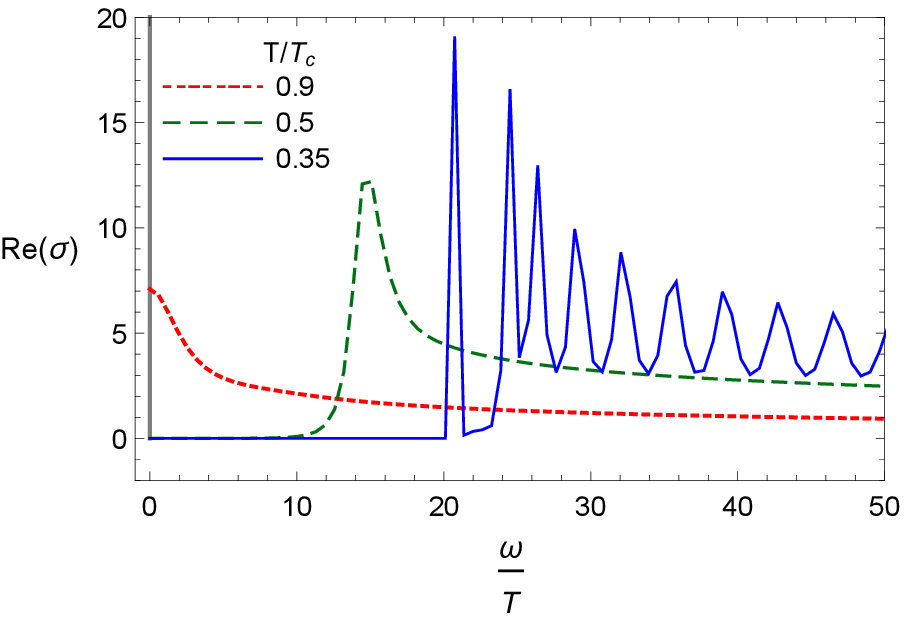}\qquad}
        \end{center}\end{minipage} \hskip+0cm
            \begin{minipage}[b]{0.325\textwidth}\begin{center}
            \subfigure[$~ d=6,~s=1$]{
    \label{fig55e}\includegraphics[width=\textwidth]{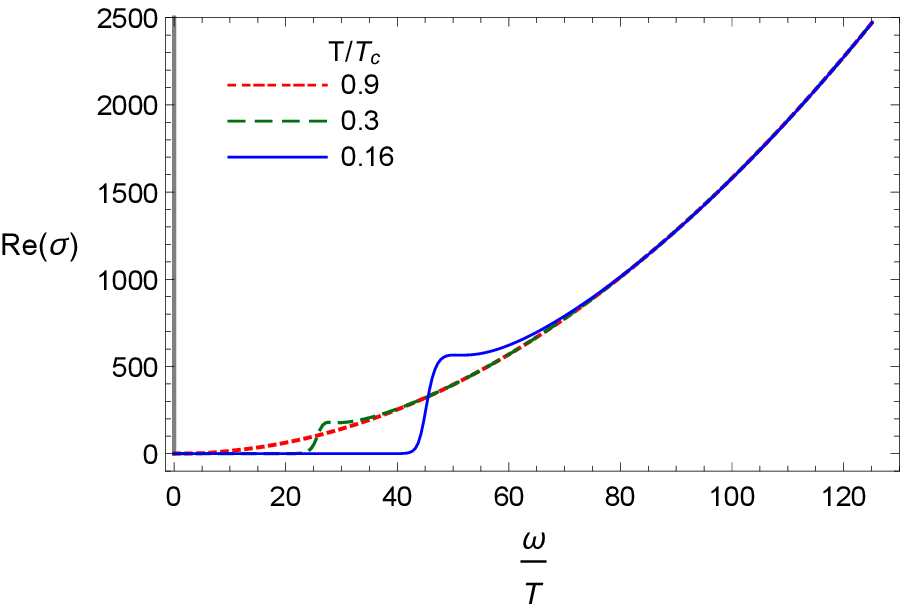}\qquad}
        \end{center}\end{minipage} \hskip0cm
        \begin{minipage}[b]{0.325\textwidth}\begin{center}
        \subfigure[$~ d=6,~s=6/4$]{
            \label{fig5f}\includegraphics[width=\textwidth]{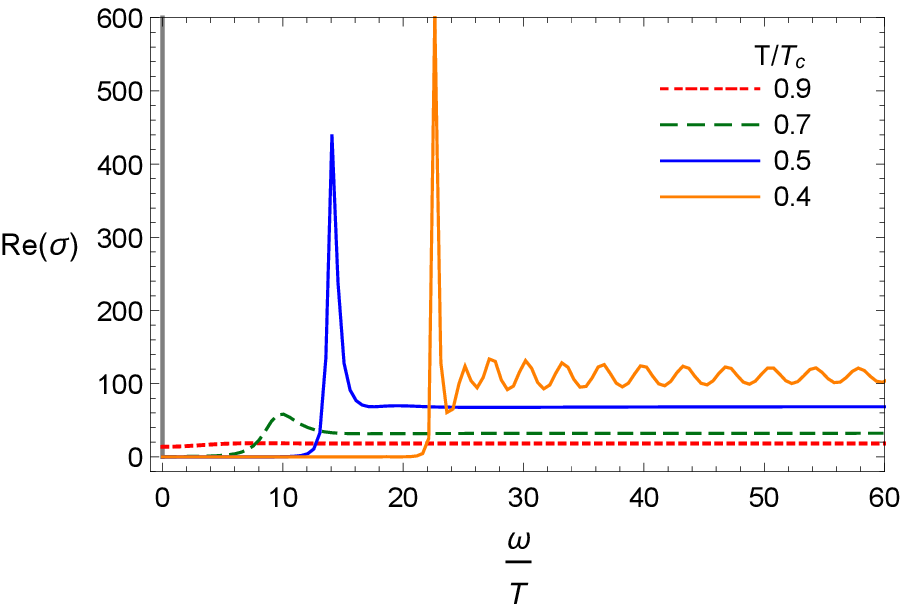}\qquad}
           \end{center}\end{minipage} \hskip+0cm
        \begin{minipage}[b]{0.325\textwidth}\begin{center}
            \subfigure[$~ d=6,~s=7/4$]{
                \label{fig555f}\includegraphics[width=\textwidth]{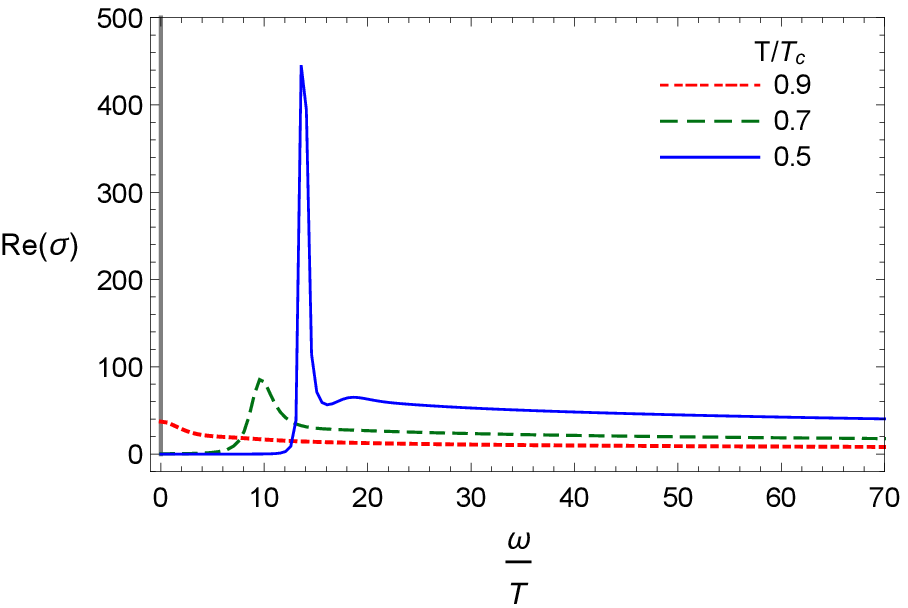}\qquad}
       \end{center}\end{minipage} \hskip+0cm
\end{center}
    \caption{The real part of conductivity for different temperature in terms of
        $\protect\omega/T$. }
    \label{fig5}
 \end{figure*}
 %%%%%%%%%%%%%%%%%%%%%%%%
  \begin{figure*}[h!]
    \begin{center}
        \begin{minipage}[b]{0.325\textwidth}\begin{center}
                \subfigure[ ~$d=4,~s=3/4$]{
                    \label{fig6a}\includegraphics[width=\textwidth]{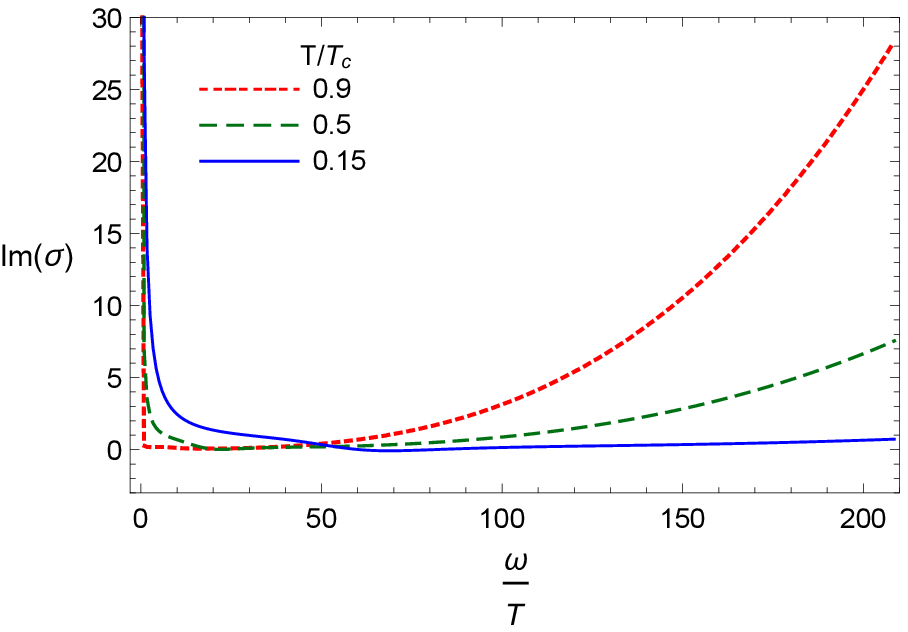}\qquad}
        \end{center}\end{minipage} \hskip+0cm
        \begin{minipage}[b]{0.325\textwidth}\begin{center}
                \subfigure[ ~$d=4, ~s=1$]{
                    \label{fig6b}\includegraphics[width=\textwidth]{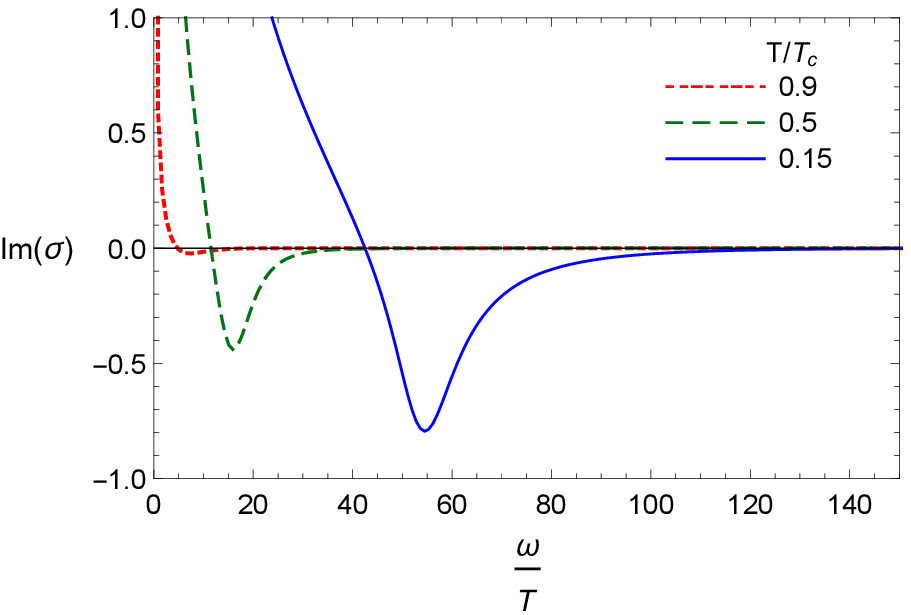}\qquad}
        \end{center}\end{minipage} \hskip0cm
        \begin{minipage}[b]{0.325\textwidth}\begin{center}
                \subfigure[ ~$d=4,~s=5/4$]{
                    \label{fig6c}\includegraphics[width=\textwidth]{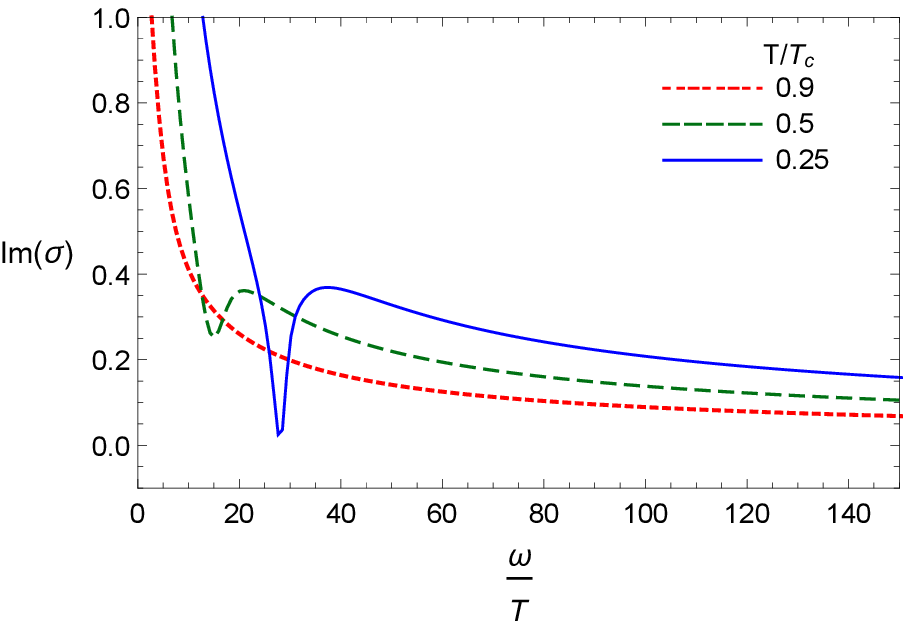}\qquad}
        \end{center}\end{minipage} \hskip0cm
        \begin{minipage}[b]{0.325\textwidth}\begin{center}
                \subfigure[$~ d=5,~s=1$]{
                    \label{fig6d}\includegraphics[width=\textwidth]{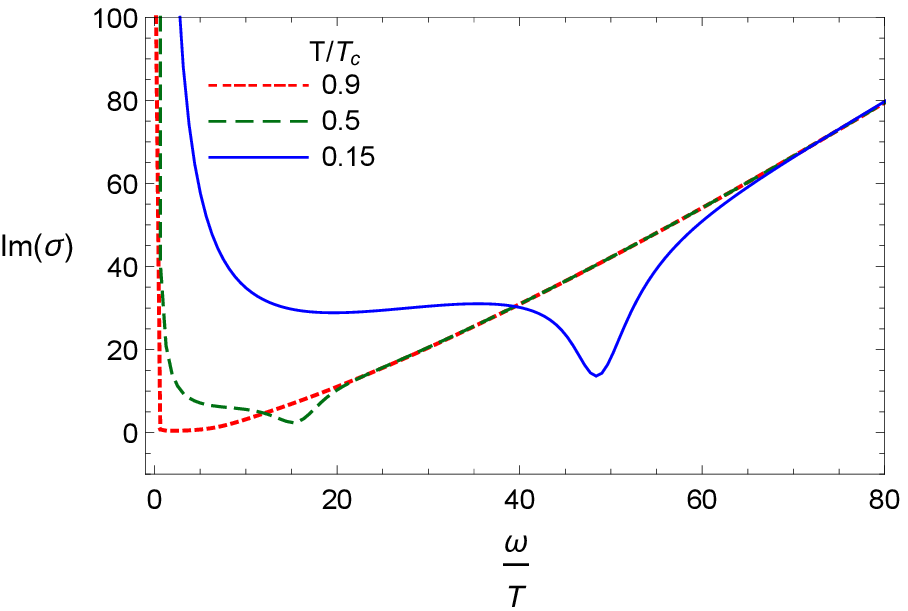}\qquad}
        \end{center}\end{minipage} \hskip0cm
        \begin{minipage}[b]{0.325\textwidth}\begin{center}
                \subfigure[$~ d=5,~s=5/4$]{
                    \label{fig6e}\includegraphics[width=\textwidth]{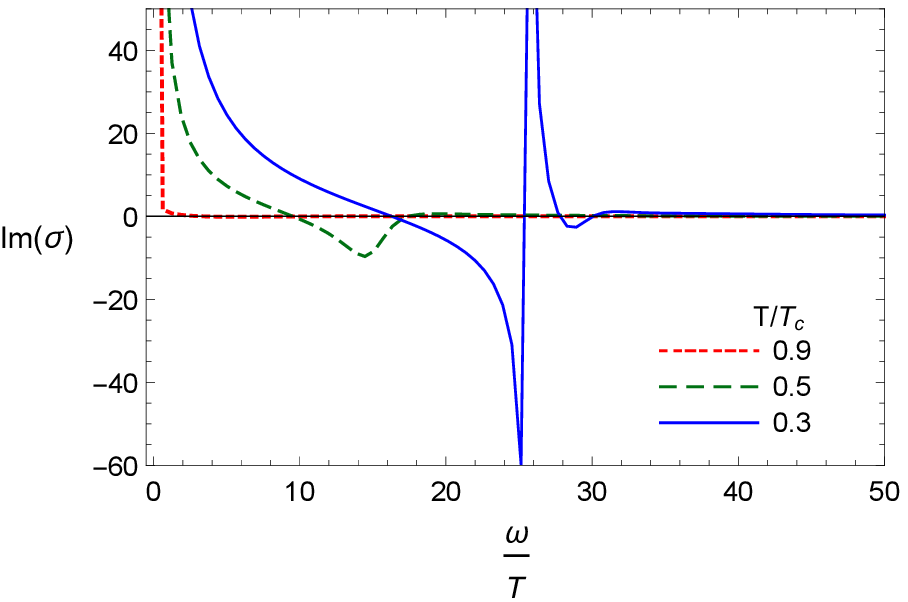}\qquad}
        \end{center}\end{minipage} \hskip0cm
        \begin{minipage}[b]{0.325\textwidth}\begin{center}
                \subfigure[$~ d=5,~s=6/4$]{
                    \label{fig6f}\includegraphics[width=\textwidth]{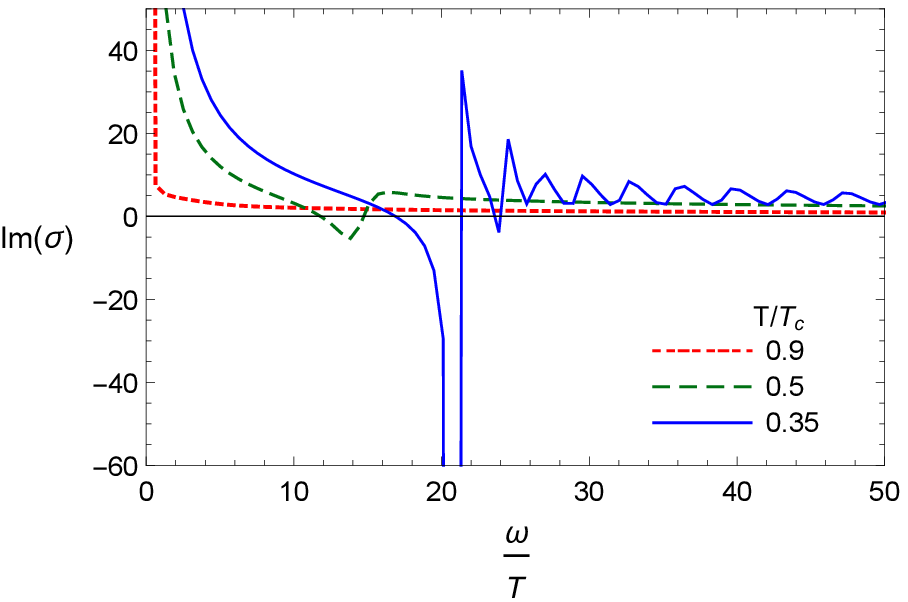}\qquad}
        \end{center}\end{minipage} \hskip+0cm
        \begin{minipage}[b]{0.325\textwidth}\begin{center}
                \subfigure[$~ d=6,~s=1$]{
                    \label{fig6e}\includegraphics[width=\textwidth]{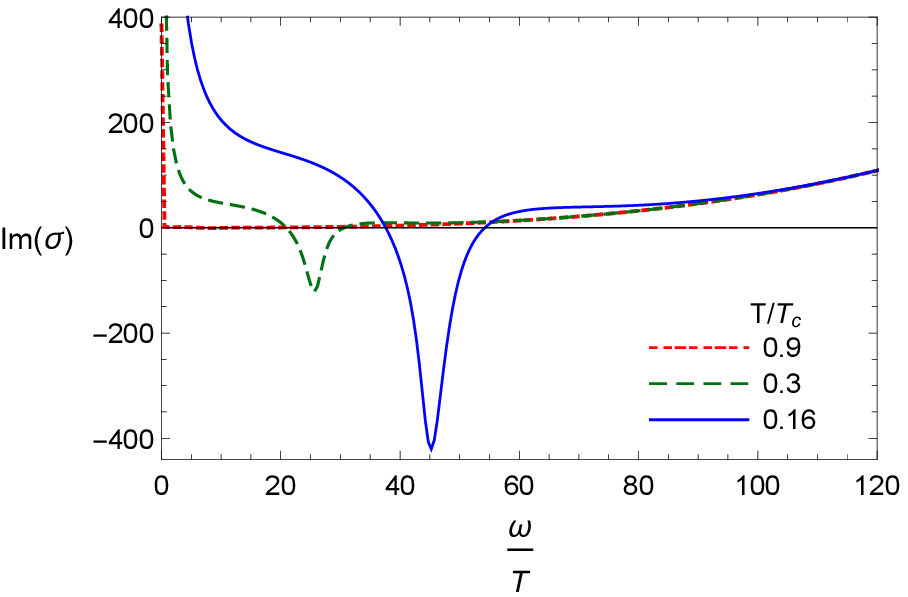}\qquad}
        \end{center}\end{minipage} \hskip0cm
        \begin{minipage}[b]{0.325\textwidth}\begin{center}
                \subfigure[$~ d=6,~s=6/4$]{
                    \label{fig6f}\includegraphics[width=\textwidth]{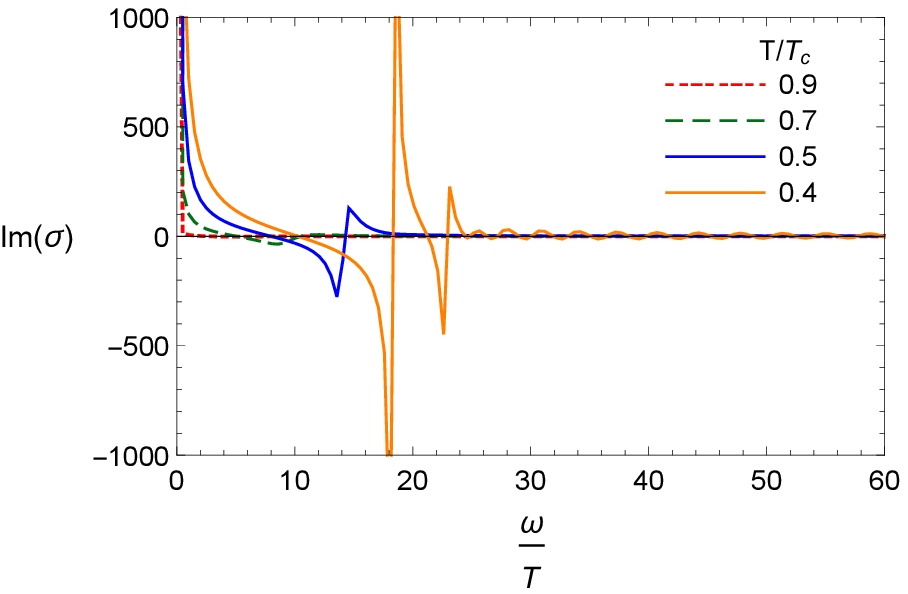}\qquad}
        \end{center}\end{minipage} \hskip+0cm
        \begin{minipage}[b]{0.325\textwidth}\begin{center}
                \subfigure[$~ d=6,~s=7/4$]{
                    \label{fig6f}\includegraphics[width=\textwidth]{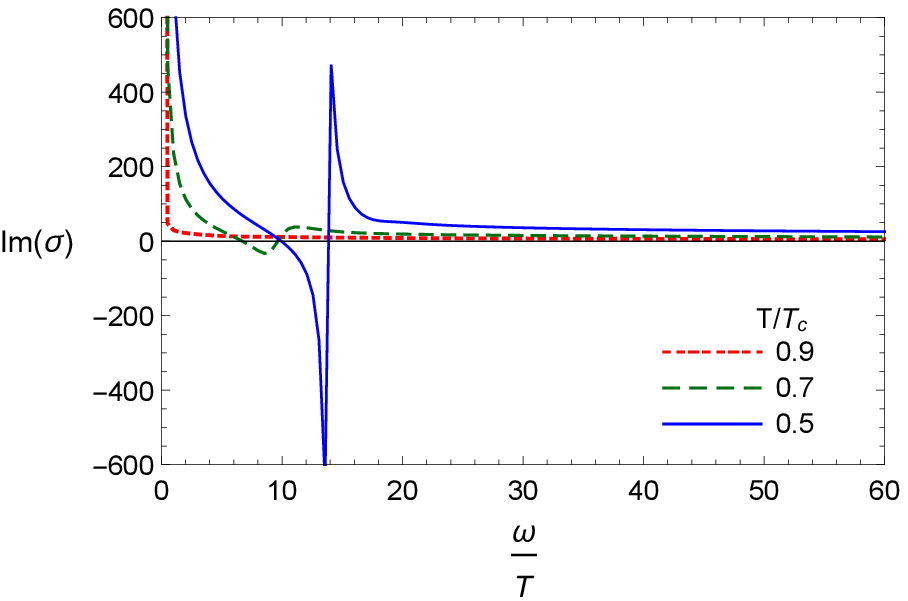}\qquad}
        \end{center}\end{minipage} \hskip+0cm
    \end{center}
    \caption{The imaginary part of conductivity for different temperature in terms of
        $\protect\omega/T$. }
    \label{fig6}
 \end{figure*}
%%%%%%%%%%%%%%%%%%%%%%%%%%%%%%%%%%%%%%%%%%%%%%%%%%%%%%%%%%%%%%%%%%%%%%%%%%%%%%%%%%
It is also interesting to study the behavior of conductivity
$\sigma$ for different values of $\omega$, and investigate the
behavior of DC and AC conductivity for the holographic
superconductor in the presence of power-Maxwell gauge field. The
behavior of the real and imaginary parts of conductivity as a
function of frequency for various power parameter $s$ and in
various dimension, at different temperature, are depicted in Figs.
\ref{fig5} and \ref{fig6}. The first point that comes out from
these plots, is the existence of a pole function in the imaginary
parts of the conductivity at $\omega=0$ for all values of $s$ and
$d$. This yields a delta function at $\omega=0$ for real part of
conductivity, which represents the infinite DC conductivity for
the holographic superconductor. For other value of $\omega$, we
see that the superconducting gap appears and becomes deep with
decreasing the temperature, which means the $\omega_{g}$ becomes
larger. At enough large frequency, the conductivity for a fixed
dimension of spacetime $d$, behaves as follows. For the conformal
invariance Power-Maxwell Lagrangian ($s=d/4$), the conductivity
does not depend on the frequency $\omega$ and tends toward a
constant value for large frequency (see Figs. \ref{fig5b},
\ref{fig5e} and \ref{fig5f}). At any dimension, for cases $s<d/4$,
at large $\omega$, the conductivity increases with increasing
$\omega$ (see Figs. \ref{fig5a}, \ref{fig5d} and \ref{fig55e}),
and for $s>d/4$, the conductivity decreases with increasing
$\omega$ (see Figs. \ref{fig5c}, \ref{fig55f} and \ref{fig555f}).
The fact that in these frequencies, the conductivity dose not
change and is a constant for the cases $s=d/4$, confirms that we
have a scale invariant theory  for these values of $s$
\cite{D.Tong}. In other words, our results is consistent with the
conformal field theory which state that the conductivity should
reach a constant value. Interestingly enough, the conductivity for
this holographic superconductor, at some dimensions and for some
values of the power parameter, has an irregular and oscillating
behavior at low temperatures, which are shown in Figs.\ref{fig55f}
and \ref{fig5f}. The frequency gap is a distinct characteristic of
a superconductor,  and we are interested in comparing the effects
of different values of the power parameter on the behavior of the
conductivity and the gap frequency $\omega_{g}$. For this purpose
we plot the conductivity in terms of $\omega/T_c$ at fixed
temperature, but for different values of the power parameter $s$.
The results are displayed in Figs. \ref{fig8} and \ref{fig9}. In
\cite{Gary T.H}, it was claimed that for the holographic
superconductor the relation $\omega_{g}\approx8 T_{c}$ is
universal. However, it can be observed from the plots that by
changing the dimension and power parameter, the frequency gap
changes slightly.

 \begin{figure}[H]
    \begin{center}
        \begin{minipage}[b]{0.325\textwidth}\begin{center}
                \subfigure[$~ d=4,~T\approx0.3 T_{c}$]{
                    \label{fig8a}\includegraphics[width=\textwidth]{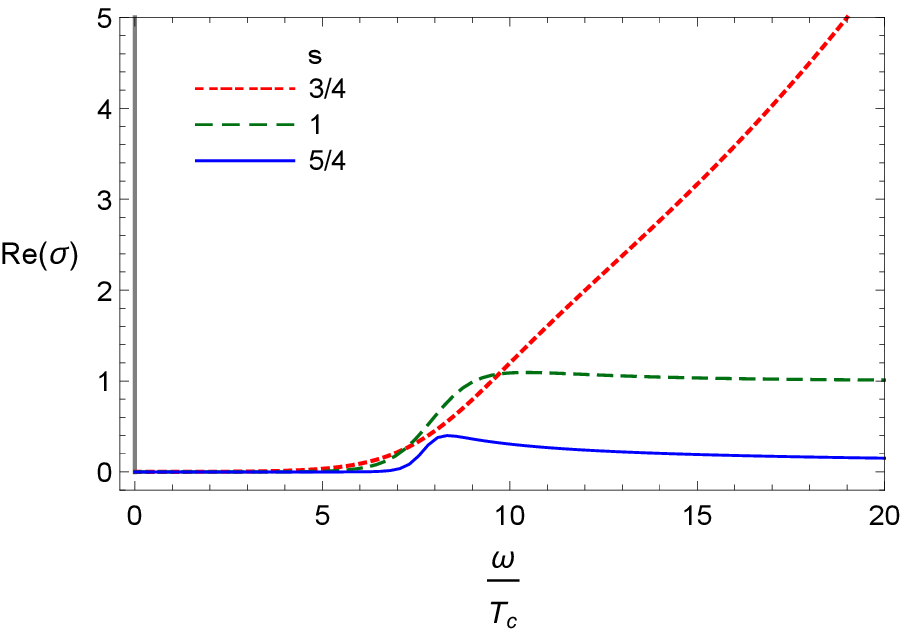}\qquad}
        \end{center}\end{minipage} \hskip+0cm
        \begin{minipage}[b]{0.325\textwidth}\begin{center}
                \subfigure[$~ d=5,~T/T_{c}\approx0.3 T_{c}$]{
                    \label{fig8b}\includegraphics[width=\textwidth]{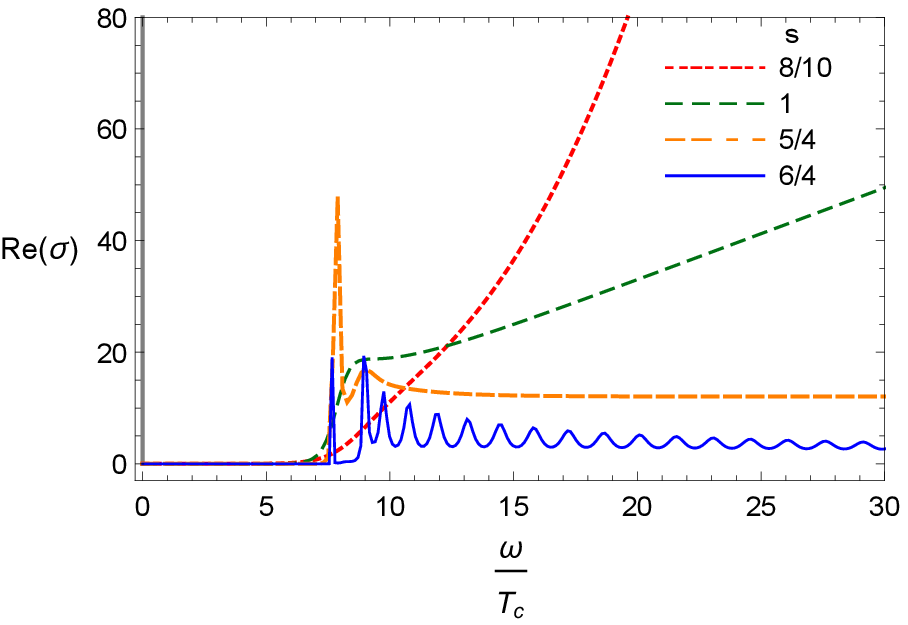}\qquad}
        \end{center}\end{minipage} \hskip0cm
        \begin{minipage}[b]{0.325\textwidth}\begin{center}
                \subfigure[$~ d=6,~T/T_{c}\approx0.5 T_{c}$]{
                    \label{fig8c}\includegraphics[width=\textwidth]{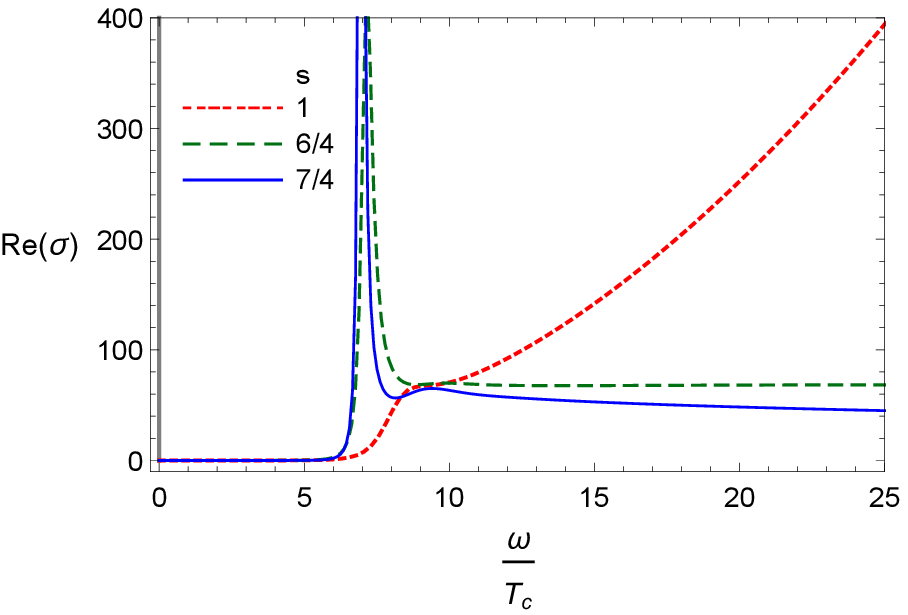}\qquad}
        \end{center}\end{minipage} \hskip0cm
    \end{center}
    \caption{The real part of conductivity for different $s$ in terms of $%
        \protect\omega/T_c $.}
    \label{fig8}
 \end{figure}

 \begin{figure}[H]
    \begin{center}
        \begin{minipage}[b]{0.325\textwidth}\begin{center}
                \subfigure[$~ d=4,~T\approx0.3 T_{c}$]{
                    \label{fig9a}\includegraphics[width=\textwidth]{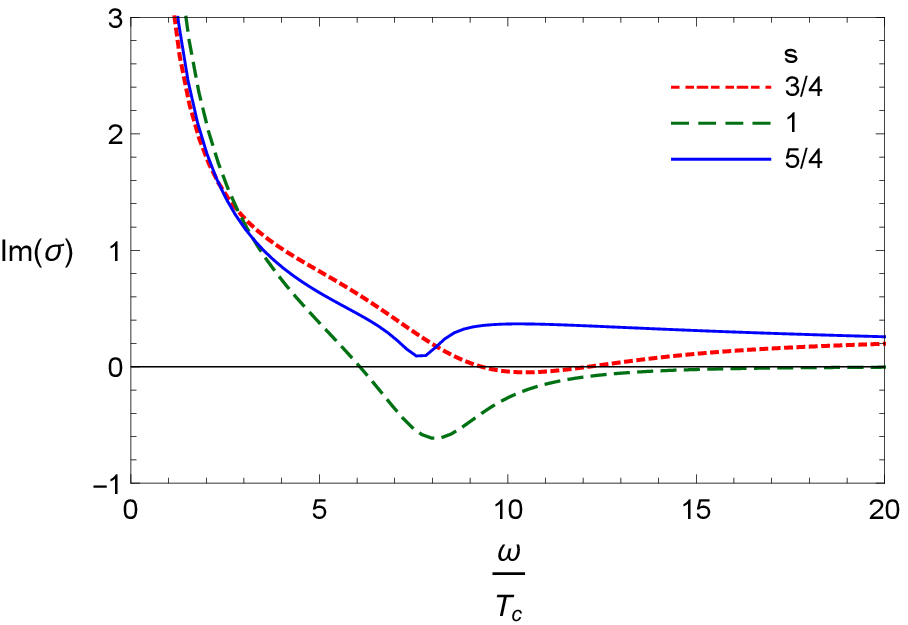}\qquad}
        \end{center}\end{minipage} \hskip+0cm
        \begin{minipage}[b]{0.325\textwidth}\begin{center}
                \subfigure[$~ d=5,~T/T_{c}\approx0.3 T_{c}$]{
                    \label{fig9b}\includegraphics[width=\textwidth]{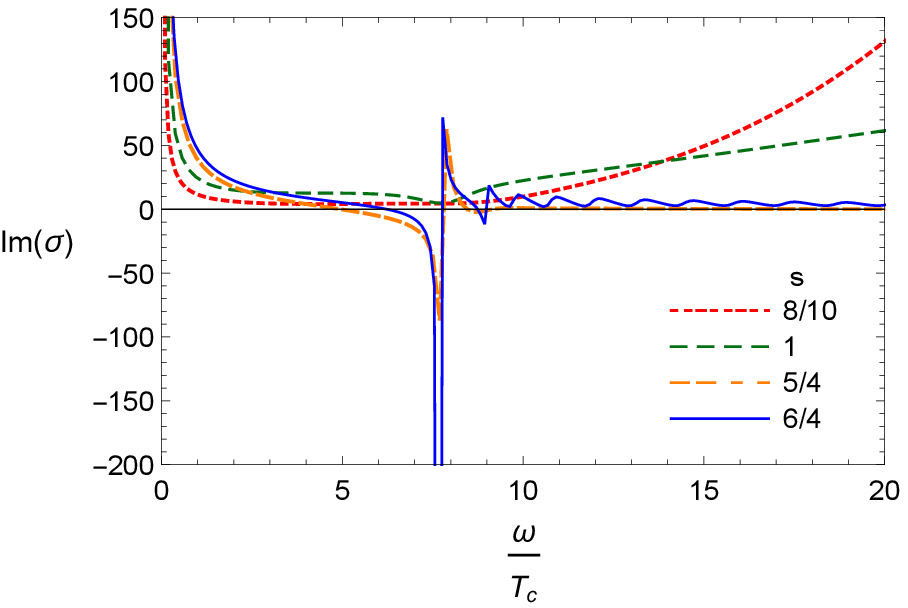}\qquad}
        \end{center}\end{minipage} \hskip0cm
        \begin{minipage}[b]{0.325\textwidth}\begin{center}
                \subfigure[$~ d=6,~T/T_{c}\approx0.5 T_{c}$]{
                    \label{fig9c}\includegraphics[width=\textwidth]{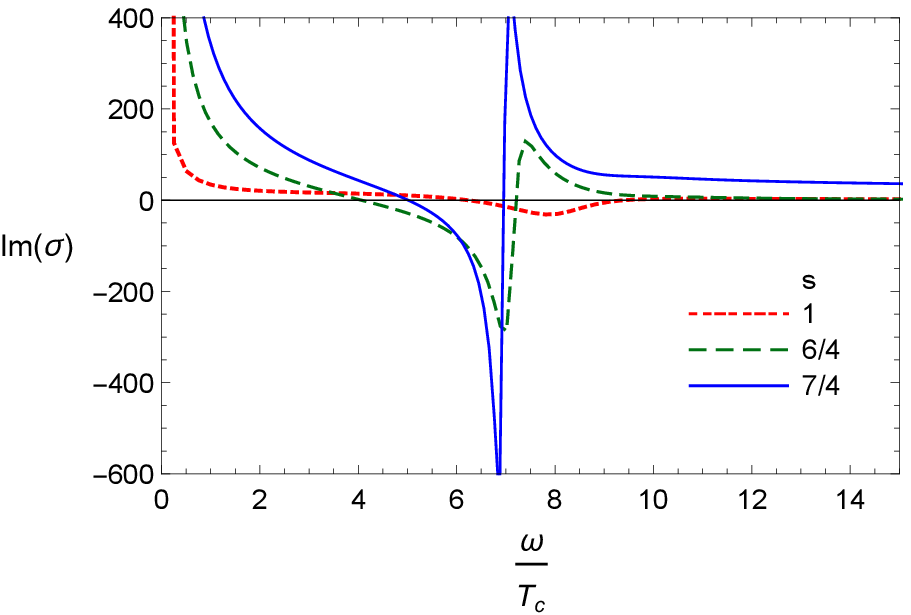}\qquad}
        \end{center}\end{minipage} \hskip0cm
    \end{center}
    \caption{The imagianry part of conductivity for different $s$ in terms of $%
        \protect\omega $.}
    \label{fig9}
 \end{figure}
%%%%%%%%%%%%%%%%%%%%%%%%%%%%%%%%%%%%%%%%%%%%%%%%%%%%%%%%%%%%%%%%%%%%%%%%%%%%%%%%%%%%%%%%%%%%
\section{The superconductor in a finite magnetic field}\label{magnetic}
It is well-known that the Meissner effect is a general feature of
the superconductors, that causes magnetic fields to be expelled
from the superconducting region. This means that the
superconductor performs a work for expelling the applied magnetic
field from the volume, the amount of required work, is
proportional to the magnetic field value. We know that the
electromagnetic free energy in a superconductor, is minimized and
the difference in the free energies between the normal and
superconducting phase can provide the energy for this work. If the
applied magnetic field is so strong and the difference between the
free energies is not enough for expelling the field from the
superconductor volume, in this case, the superconductor will exit
the superconducting phase and enter the normal phase.  At any
fixed temperature, the maximum values of the external magnetic
field for which above of it, the superconductor cannot have
condensation, called as the critical field \cite{Har2}.  Here, we
shall investigate the effects of an external magnetic field on the
holographic superconductor and would like to find the effect of
electrodynamics on the process of expelling the magnetic field. At
first, we immerse the superconductor (that is in the normal phase)
into an external magnetic field, and then we decrease the
temperature until the condensation occurs at a particular
temperature. This temperature is the critical temperature for that
value of the magnetic field, and when we change the value of
applied magnetic field, this critical temperature changes, too.
The field equations, are the same as in Eqs. (\ref{enequation}),
(\ref{gaugefield}) and (\ref{scalarfield}), with the difference
that in this section we add a magnetic component to the Maxwell
field. We consider
 \begin{eqnarray}\label{Amag}
A_{a}dx^{a}=\phi(r,x)~dt+A_{y}(r,x) dy,~~~~  \   \psi=\psi(r,x).
\end{eqnarray}
We rewrite the field equations by considering these assumptions.
The original equations in which, the power parameter is a
variable, are too long and for the economic reasons we do not
present it here. When $s=1$, the equations reduce to the Maxwell
form \cite{Har2},
 \begin{eqnarray}\label{phimag}
\frac{\partial ^{2}}{\partial r^{2}}\phi(r,x)+\frac{1}{r^{2}~f(r)}
\frac{\partial^{2}}{\partial x^2} \phi(r,x)-\frac{(d-2)}{r}
\frac{\partial}{\partial r} \phi(r,x)-\frac{2\phi(r,x)
\psi(r,x)^{2}}{f(r)}=0,
 \end{eqnarray}
\begin{eqnarray}\label{Aymag}
&&\frac{\partial ^{2}}{\partial
r^{2}}A_{y}(r,x)+\frac{1}{r^{2}~f(r)} \frac{\partial^{2}}{\partial
x^2}A_{y}(r,x)+\frac{(d-4)}{r} \frac{\partial}{\partial r}
A_{y}(r,x)-\frac{2A_{y}(r,x) \psi(r,x)^{2}}{f(r)}\nonumber \\
&&+\frac{\partial f(r)/\partial r}{f(r)}\frac{\partial}{\partial
r} A_{y}(r,x)=0,
\end{eqnarray}
\begin{eqnarray}\label{psimag}
&&\frac{\partial ^{2}}{\partial r^{2}}\psi(r,x)
+\frac{1}{r^{2}~f(r)} \frac{\partial^{2}}{\partial x^2}
\psi(r,x)+\frac{(d-2)}{r} \frac{\partial}{\partial r}
\psi(r,x)-\frac{m^{2}\psi(r,x)}{f(r)}\nonumber \\
&&+\frac{\phi(r,x)^{2}\psi(r,x)}{f(r)^2}-\frac{A_{y}(r,x)^{2}}{r^{2}f(r)}+\frac{\partial
f(r)/\partial r}{f(r)}\frac{\partial}{\partial r} \psi(r,x)=0.
\end{eqnarray}
In the following we study the asymptotic behaviour of the field
equations. We further assume the scalar field $\psi$ is much
smaller than the magnetic field. This is true near the critical
temperature when the condensation is formed. Therefore, we treat
the scalar field as a perturbation in the bulk and obtain the
solution for the equations \cite{albash}. In this limit, $\psi$ in
Eqs. (\ref{phimag}),(\ref{Aymag}) vanishes, and it is easy to find
the solutions. Also, the original versions of these equations with
variable $s$ also becomes very simpler by employing perturbative
limit. Moreover, we suppose the applied magnetic field has a
constant value at the boundary and do not respect the radial
coordinate $r$, and $\phi=\phi(r)$. Given these conditions, we
find an appropriate solution for the field $A_{y}$ which is
$A_{y}=Bx$, corresponding to a uniform magnetic field, when $B$ is
a constant, that is dual to magnetic field at the boundary. Since
our purpose here is to investigate the effects of magnetic field,
in the general case. Therefore, we consider the backreaction of
the applied magnetic field on the background geometry. In other
words, we would like to consider a fully back-reacted magnetically
charged black hole in the bulk and neglect the backreaction of the
scalar field $\psi$ and gauge field $\phi$. Therefore, by solving
Eq. (\ref{enequation}), we find
\begin{equation}
f(r) =\left\{
\begin{array}{ll}
r^{2}-\frac{(4r_{+}^{4}+B^{2})}{4rr_{+}}+\frac{B^2}{4r^{2}}\ , & \ \ ~~\mathrm{for}\ d=4,\\
&  \\
r^{2}-\frac{r_{+}^4}{r^2}+\frac{B^{2}}{6r^{2}}~\ln(\frac{r_{+}}{r}),
& \ \quad
\mathrm{for}\ d=5,\\
&  \\
r^{2}-\frac{r_{+}^{5}}{r^3}+\frac{ B^{2}r_{+}}{8 r^3}-\frac{B^2}{8r^2}, & \ \ ~~\mathrm{for}\ d=6.\\
&  \\
\end{array}%
\right.  \label{f(r)}
\end{equation}%
As before, from the definition of the Hawking temperature
$T=\frac{f^{\prime}(r_{+})}{4\pi}$, we find the temperature of
this superconductor in the presence of the magnetic field $B$,
\begin{equation}
T =\left\{
\begin{array}{ll}
\frac{12-B^2}{16\pi}\ , & \ \ ~~\mathrm{for}\ d=4,\\
&  \\
\frac{24-B^2}{24\pi}, & \ \quad
\mathrm{for}\ d=5,\\
&  \\
\frac{40-B^2}{32\pi}, & \ \ ~~\mathrm{for}\ d=6.\\
&  \\
\end{array}%
\right.  \label{TT}
\end{equation}%
Inserting $B=0$ in the functions $f(r)$ and $T$, one can obtain
the metric function and the Hawking temperature for the higher
dimensional Schwarzschild-AdS black holes \cite{Har1}. With
$A_{y}$ and $f(r)$ at hand, one can substitute them in the
equation of $\phi$ and $\psi$ and solve these equation
numerically.  But before that, the equation for
$\psi(r,x)=X(x)R(r)$ can be separated, yielding the following
differential equations
  \begin{eqnarray}\label{Rmag}
R^{\prime\prime}(r) +\left(\frac{f^{\prime
}}{f}+\frac{d-2}{r}\right) R^{\prime }(r) +\bigg(\frac
{\phi^2(r)}{f^2(r)}-\frac{m^2}{f(r)}\bigg) R(r)=\lambda R(r),
  \end{eqnarray}
\begin{eqnarray}\label{Xmag}
X^{\prime \prime}(x)-B^{2} x^{2} X(x)=-\lambda X(x).
\end{eqnarray}
when $\lambda$ is the separation constant $\lambda=qnB$ with
$n=1$.  The solution for $X(x)$, has been discussed in Refs.
\cite{Har2,albash}.

\begin{figure}[H]
    \begin{center}
        \begin{minipage}[b]{0.325\textwidth}\begin{center}
                \subfigure[$~ d=4,s=1,B=1$]{
                    \label{fig11a}\includegraphics[width=\textwidth]{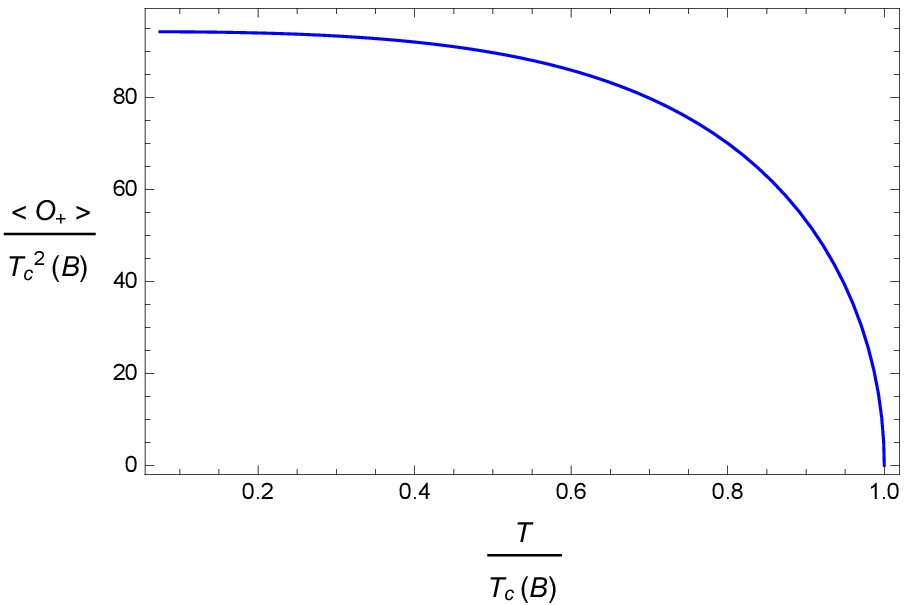}\qquad}
        \end{center}\end{minipage} \hskip+0cm
        \begin{minipage}[b]{0.325\textwidth}\begin{center}
                \subfigure[$~ d=5,s=5/4,B=1$]{
                    \label{fig11b}\includegraphics[width=\textwidth]{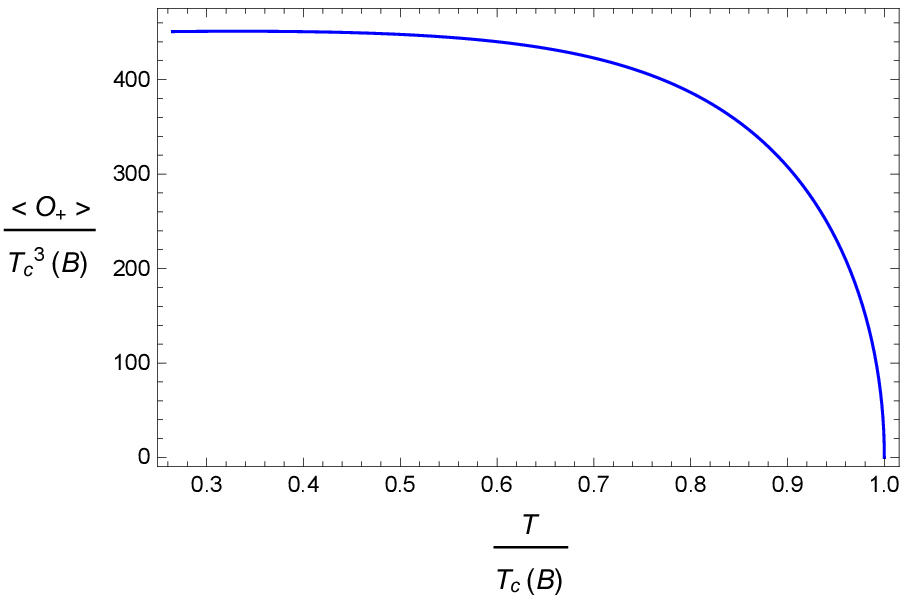}\qquad}
        \end{center}\end{minipage} \hskip0cm
        \begin{minipage}[b]{0.325\textwidth}\begin{center}
                \subfigure[$~ d=6,s=6/4,B=1$]{
                    \label{fig11c}\includegraphics[width=\textwidth]{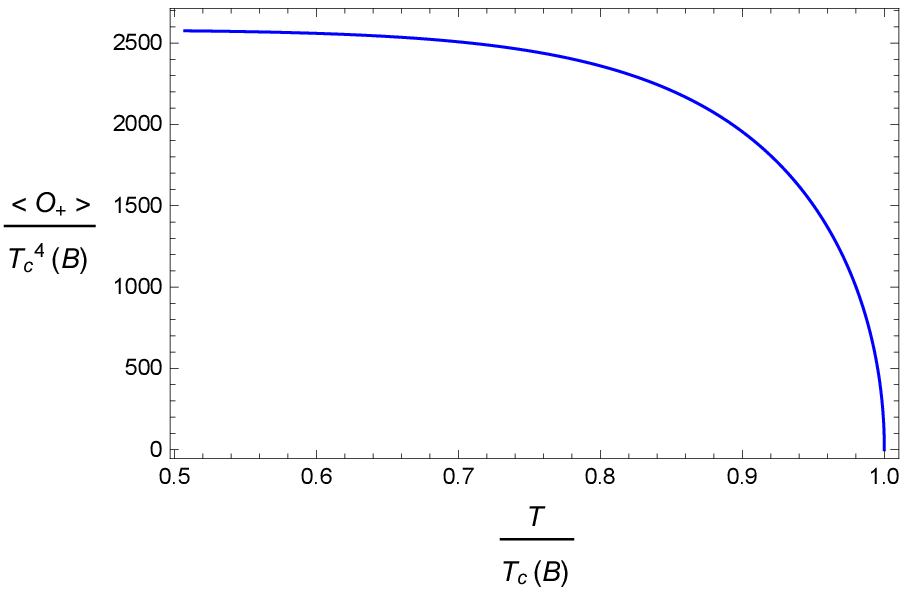}\qquad}
        \end{center}\end{minipage} \hskip0cm
    \end{center}
    \caption{The dimensionless condensation in the presence of external magnetic field versus temperature.}
    \label{fig11}
\end{figure}

\begin{figure}[H]
    \begin{center}
        \begin{minipage}[b]{0.325\textwidth}\begin{center}
                \subfigure[$~ d=4$]{
                    \label{fig10a}\includegraphics[width=\textwidth]{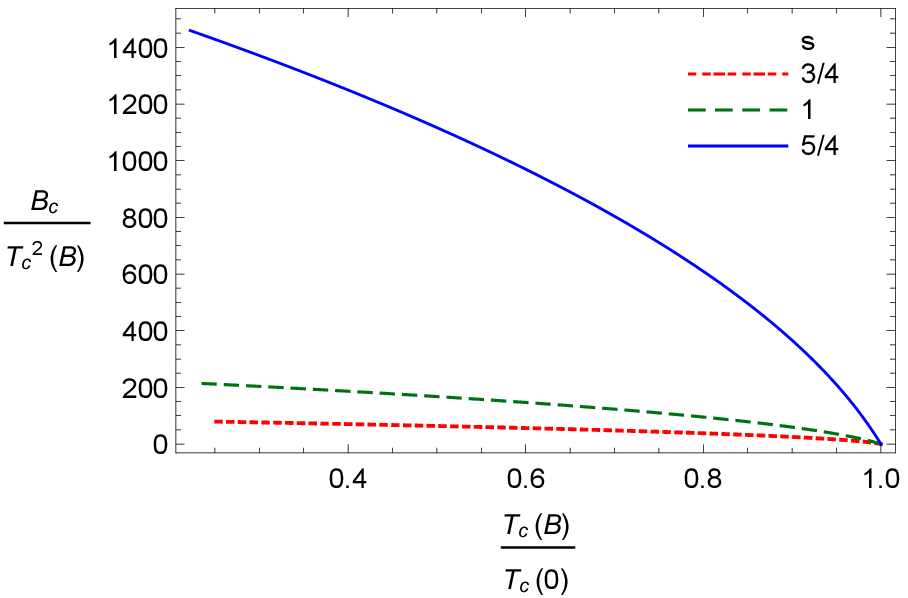}\qquad}
        \end{center}\end{minipage} \hskip+0cm
        \begin{minipage}[b]{0.325\textwidth}\begin{center}
                \subfigure[$~ d=5$]{
                    \label{fig10b}\includegraphics[width=\textwidth]{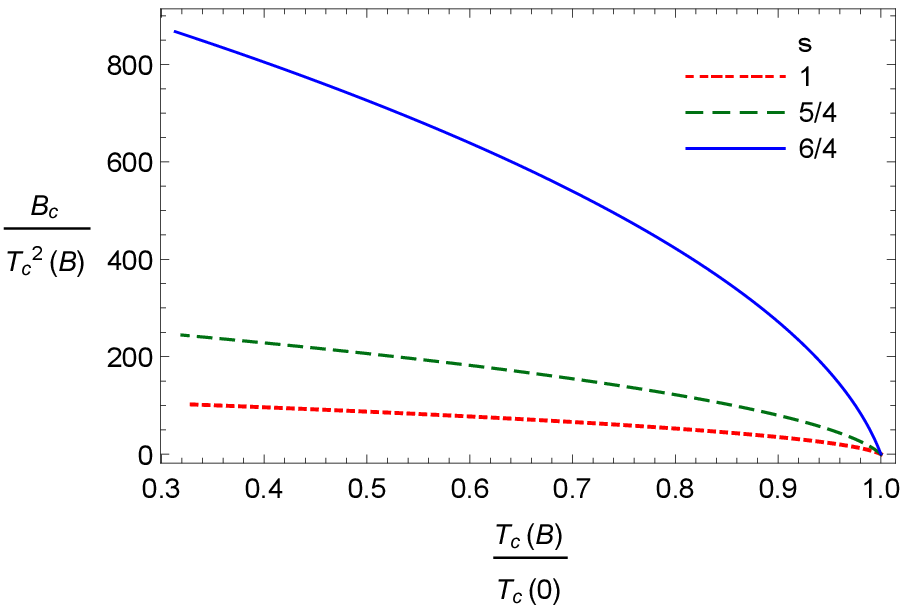}\qquad}
        \end{center}\end{minipage} \hskip0cm
        \begin{minipage}[b]{0.325\textwidth}\begin{center}
                \subfigure[$~ d=6$]{
                    \label{fig10c}\includegraphics[width=\textwidth]{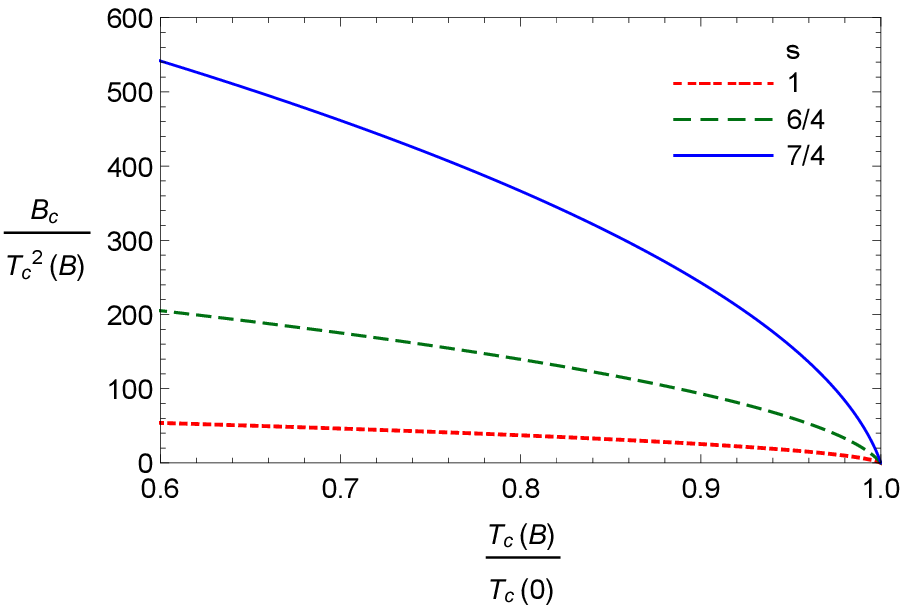}\qquad}
        \end{center}\end{minipage} \hskip0cm
    \end{center}
    \caption{The critical magnetic field versus temperature.}
    \label{fig10}
\end{figure}

In order to check the condensation for some values of the applied
magnetic field, we solve the field equations for $\phi$ and $\psi
$ in the bulk.  We expect that the condensation starting point,
should be dependent on the values of the magnetic field $B$.
Fig.\ref{fig11}, represents the dimensionless condensation with
respect to dimensionless temperature, for the conformal invariance
case and for a fixed value of the magnetic field $B$. We can see
that the dimensionless condensation in the presence of the
magnetic field has the same behavior that had in the absence of a
magnetic field, but it is clear that the value of this quantity
has changed.
Use the results to achieve the critical temperature $T_{c}(B)$.
Now we can produce a curve of solutions that have plotted in
Fig.\ref{fig10}. It is
interesting to plot the dimensionless parameters too. in the
presence of power-Maxwell electrodynamics, and for all values of
$d$, we find the dimension of $B$, always two.
%%%%%%%%%%%%%%
$T_{c}(0)$ is the critical temperature in the absence of the
magnetic field, and $T_{c}(B)$ is the critical temperature in the
presence of the magnetic field $B$. In the region that is below
the plots, there is superconducting condensation. Actually, the
below region, displays all values of temperature and
simultaneously, the values of the critical magnetic field for
which the superconductor remains in the superconducting phase. As
grow up the temperature or the applied magnetic field, depending
on the spacetime dimension and the power parameter, the
superconductor leave the superconducting phase at a point (on the
lines of  the plots). We can also see that for all values of $s$
and $d$, as the magnetic field increases, the critical temperature
decreases, so the condensation becomes harder. This is  an
expected result, which can be interpreted as the Meissner-like
effect for holographic superconductors.
%%%%%%%%%%%%%%%%%%%%%%%%%%%%%%%%%%%%%%%%%%%%%%%%%%%%%%%%%%%%%%%%%%%%%%%
\section{Closing Remarks}
In this paper, based on the numerical shooting method, we have
explored the holographic $s$-wave superconductor in the presence
of power-Maxwell electrodynamics and in all higher dimensions. We
have derived the critical temperature for this kind of
superconductor and found out that the it decreases with increasing
the power parameter $s$, which means it is harder for the
condensation to form in the presence of stronger electrodynamics.
We observed that when the power parameter is smaller that one
($s<1$) the critical temperature increases as well \cite{SSM}. In
this case the strength of the gauge field decreases for
power-Maxwell case comparing to the Maxwell case ($s=1$).
Therefore, the effects of the gauge field on the background
decrease as well. This implies that it is quite reasonable to take
still the probe limit when the gauge field is in the form of the
power-Maxwell electrodynamics. The obtained results are also
consistent with the fact that in higher dimensional systems, the
phase transition are easier to achieve. We have obtained the
condensation value and the critical exponent and showed that for
all values of the power parameter $s$ and dimension $d$, the
critical exponent is $1/2$. We observed that in higher dimensions
and for larger value of $s$, the re-scaled condensation operator
increase, explicitly. We explored the electrical conductivity by
applying an external electric field to the system. In this
regards, we obtained infinite DC conductivity for this
superconductor and showed that the superconducting gap appeared
and becomes deep with decreasing the temperature. We observed that
at enough large frequency, for a fixed dimension of spacetime $d$,
and for the conformal invariance power-Maxwell Lagrangian $s=d/4$,
the conductivity does not depend on the frequency $\omega$ and
tends toward a constant value for large frequency. This is
consistent with the fact that in conformal field theory, the
conductivity reaches a constant value at large frequency. We also
investigated the cases with  $s<d/4$ and $s>d/4$, separately. In
the former case, the conductivity increases with increasing
$\omega$, while in the latter one the conductivity decreases with
increasing $\omega$. Also, the ratio $\omega_{g}/T_{c}$ varies
slightly with changing the dimension and power parameter. We have
also immersed the superconductor into an external magnetic field
and studied the behavior of the critical magnetic field. We found
out that for all value of $d$ and $s$, the magnetic field has an
expected behavior. As the magnetic field increases, the critical
temperature decreases, so the condensation becomes harder. This
can be interpreted as the Meissner-like effect for holographic
superconductors.
%%%%%%%%%%%%%%%%%%%%%%%%%%%%%%%%%%%%%%%%%%%%%%%%%%%%%%%%%%%%%%%%%%%%%%%%%%%%%%%%%%%
\section*{Acknowledgments}
We thank the referee for constructive comments which helped us
improve our paper. We  also thank Shiraz University Research
Council. The work of AS has been supported financially by Research
Institute for Astronomy and Astrophysics of Maragha (RIAAM), Iran.
%%%%%%%%%%%%%%%%%%%%%%%%%%%%%%%%%%%%%%%%%%%%%%%%%%%%%%%%%%%%%%%%%%%%%%%%%%%%%%%%%%%%

\end{document}